\documentclass[12pt]{article}

\usepackage{array}
\usepackage{epsfig}
\usepackage{amssymb}
\usepackage{graphics,graphpap}

\renewcommand{\thefootnote}{\fnsymbol{footnote}}

\newcommand{\quark}{\langle \bar q q\rangle}
\newcommand{\mixed}{\langle \bar q \sigma gG q\rangle}
\newcommand{\squark}{\langle \bar s s\rangle}

\newcommand{\smixed}{\langle \bar s \sigma gG s\rangle}

\newcommand{\gluon}{\left\langle \frac{\alpha_s}{\pi}\,G^2\right\rangle}
\newcommand{\bra}{\langle}
\newcommand{\ket}{\rangle}
\newcommand{\ds}{\displaystyle}
\newcommand{\wt}{\widetilde}

\def\s#1{\setbox0=\hbox{$#1$}%
  \rlap{\ifdim\wd0>.7em\kern.22\wd0\else\kern.1\wd0\fi /}#1}

\makeatletter
\def\slash#1{\rlap/{\mkern-2mu {#1}}} 
\makeatother

\begin{document}

\begin{titlepage}
\begin{flushright}
\begin{tabular}{l}
IPPP/06/01\\
DCPT/06/02\\
CERN--PH--TH/2006--026
\end{tabular}
\end{flushright}
\vskip1.5cm
\begin{center}
{\Large \bf \boldmath
Higher-Twist Distribution Amplitudes\\[7pt] of the K Meson in QCD}
\vskip1.3cm 
{\sc
Patricia Ball\footnote{Patricia.Ball@durham.ac.uk}$^{,1,2}$,
V.M.~Braun\footnote{Vladimir.Braun@physik.uni-regensburg.de}$^{,3}$ and 
A. Lenz\footnote{Alexander.Lenz@physik.uni-regensburg.de}$^{,3}$}
  \vskip0.5cm
        $^1$ IPPP, Department of Physics,
University of Durham, Durham DH1 3LE, UK \\
\vskip0.4cm
$^2$ CERN, CH--1211 Geneva 23, Switzerland\\
\vskip0.4cm
$^3$ Institut f\"ur Theoretische Physik, \\ Universit\"at Regensburg,
D--93040 Regensburg, Germany

\vskip2cm


\vskip3cm

{\large\bf Abstract\\[10pt]} \parbox[t]{\textwidth}{
We present a systematic study of twist-3 and twist-4 light-cone distribution 
amplitudes of the $K$ meson in QCD. 
The structure of $SU(3)$-breaking corrections is studied in
detail. Non-perturbative input parameters are estimated from QCD sum
rules and renormalons. As a by-product, we give a complete reanalysis
of the twist-3 and -4 parameters of the
$\pi$-meson distribution amplitudes; some of the results differ from
those usually quoted in the literature.
}

\vfill

\end{center}
\end{titlepage}

\setcounter{footnote}{0}
\renewcommand{\thefootnote}{\arabic{footnote}}
\renewcommand{\theequation}{\arabic{section}.\arabic{equation}}

\newpage

\section{Introduction}
\setcounter{equation}{0}

The discovery of an approximate $SU(3)$ flavour symmetry of strong interactions 
\cite{Gell-Mann:1964nj,Ne'eman:1961cd} predates the quark model and has been of paramount 
importance in all subsequent developments. This symmetry has its origin in the smallness
of up, down and strange quark masses with respect to the QCD scale
$\Lambda_{\rm QCD}$.
In static hadron properties, such as masses, magnetic moments, decay constants, etc.,
it is accurate to about 1 to 3\% for quantities related by isospin and
to about 20\% for those related by U- and V-spin. 
The breaking of $SU(3)$ in dynamical observables can be larger and up to now is not fully 
understood. One particularly striking example is the weak radiative decay $\Sigma \to p\gamma$:
the experimental azimuthal asymmetry in this decay is $\alpha_\gamma = -0.76\pm 0.08$ 
\cite{PDG}, which according to Hara's theorem \cite{Hara} implies 100\% $SU(3)$
symmetry violation.
  
In recent years  $SU(3)$-symmetry-breaking effects in heavy-meson
decays have attracted increasing interest. These processes can be
treated in heavy-quark expansion, which has proved a very powerful
theoretical tool, so that in some cases, for instance weak radiative
decays, $B\to\rho\gamma$ vs.\ $B\to K^*\gamma$, the 
uncertainty in $SU(3)$ breaking is actually the dominant source of theoretical error.    
The particular challenge of such processes is that, in the presence of
a hard scale, 
hard exclusive reactions are dominated by rare configurations
of the hadrons' constituents: either only valence-quark configurations contribute and all quarks have small 
transverse separation (hard mechanism), or one of the partons carries most of the hadron momentum 
(soft or Feynman mechanism) \cite{BLreport}. The size of
$SU(3)$-breaking 
effects in such rare configurations cannot be
deduced from the symmetry breaking in static properties, where the bulk
of the wave-function contributes.

Hard contributions are simpler to treat than their soft counterparts
and their structure is well  understood, see Ref.~\cite{exclusive}.
They can be calculated in terms of the hadron distribution amplitudes (DAs) which 
%
describe
the momentum-fraction distribution of partons at zero transverse separation 
in a particular Fock state, with a fixed number of constituents. DAs are ordered by increasing twist; 
the leading-twist-2 meson DA $\phi_{2;P}$, which describes the momentum
 distribution of the valence quarks in the meson $P$, is related to the
 meson's Bethe--Salpeter wave function $\phi_{P,BS}$ by an integral
 over transverse momenta:
$$
\phi_{2;P}(u,\mu) = Z_2(\mu) \int^{|k_\perp| < \mu} \!\!d^2 k_\perp\,
\phi_{P,BS}(u,k_\perp).
$$
Here $u$ is the quark momentum fraction, $Z_2$ is the renormalization factor (in the light-cone gauge) 
for the quark-field operators in the wave function, and $\mu$ denotes the renormalization scale.
The study of the leading-twist DA of the pion has attracted much attention in
the literature, whereas the status of  $SU(3)$-breaking effects 
that are responsible for the difference between the kaon and the pion DAs has been controversial 
for a while \cite{BB03}. These corrections have been recently reconsidered 
in the framework of QCD sum rules in Refs.~\cite{KMM04,BL04,BZ05,BZ06}, 
with a consistent picture finally emerging. We will give a short review of these developments 
below.  

Higher-twist DAs are much more numerous and describe either
contributions of ``bad'' components in the wave function, or contributions
of transverse motion of quarks (antiquarks) in the leading-twist
components, or contributions of higher Fock states with additional
gluons and/or quark--antiquark pairs. Within the hard-rescattering picture, 
the corresponding contributions to the hard exclusive reactions are
suppressed by a power (or powers) of the large   
momentum $Q$, which explains why they have received less attention.

In turn,  soft contributions are intrinsically non-perturbative  and  cannot be further
reduced (or factorized) in terms of simpler quantities without additional assumptions. At present,
they can only be estimated using light-cone sum rules
\cite{BBK89,BF89,CZ90}, see Refs.~\cite{LCSR,BZ04} for sample 
applications to heavy quark decays.
In this technique soft contributions are extracted from the dispersion relations for 
suitable correlation functions, by introducing an auxiliary
``semi-hard'' scale (Borel parameter) at 
which the two different representations of the correlation function, in terms of quarks and
in terms of hadronic states, are matched. In calculations of this kind, the 
necessary non-perturbative input again reduces to DAs, and 
the higher-twist DAs play a very important role,
since they are only suppressed by powers of the Borel parameter and
not by powers of the hard scale $Q$.
The crucial point and main technical difficulty in the construction of higher-twist DAs is the 
necessity to satisfy the exact equations of motion (EOM), which yield
relations between 
physical effects 
of different origins: for example, using EOM, the contributions of
orbital angular momentum 
in the 
valence component of the wave function can be expressed (for mesons)
in terms of 
contributions of higher 
Fock states. An appropriate framework for implementing these
constraints was developed in Ref.~\cite{BF90}: it is based on the derivation 
of EOM relations for non-local light-ray operators~\cite{string},
which are solved order 
by order in 
the conformal expansion; see Ref.~\cite{BKM03} for a review and further references.   
In this way it is possible to construct self-consistent approximations for the DAs, which involve
a minimum number of hadronic parameters. Another approach, based on the study of renormalons, 
was suggested for twist 4 in Refs.~\cite{Andersen99,BGG04}: this
technique is appealing as it allows one to 
obtain an estimate of  high-order contributions to the conformal
expansion which are usually omitted.
In this paper, we generalize this approach to include $SU(3)$-breaking 
corrections and show how to combine renormalon--based estimates of ``genuine'' twist-4 effects
with meson mass corrections. 

Pion DAs of twist 3 and  4 have already been studied in
Ref.~\cite{BF90}. In Ref.~\cite{PB98}, these results
were extended to the pseudoscalar octet; they include  
those meson-mass corrections that
break chiral symmetry, while still preserving G-parity.  $SU(3)$-breaking in the 
normalization of the twist-4 DA was estimated in Ref.~\cite{KMM03}. 
In this paper we continue the analysis of twist-3 and 4 DAs of the $K$ meson and present,
for the first time, the complete set of DAs, including also G-parity-breaking terms   
that vanish in the limit of quarks with equal mass.
The results are of direct relevance to the
discussion of, for instance, $B$-meson decays into light mesons using  light-cone sum rules and 
also in the SCET framework. We refrain from an analysis of the $\eta$
DAs, as the inclusion of the singlet part is crucial for
phenomenological applications, but goes beyond
the scope of this paper, and in fact deserves a separate study.

Our paper is organized as follows: in Section~\ref{sec:2} we explain
notation and review the existing information on leading-twist
DAs. Section~\ref{sec:3} is devoted to twist-3 DAs: we give their general classification,
calculate meson-mass corrections, perform a conformal expansion and formulate models
in terms of a few non-perturbative parameters. A similar programme is carried out 
for twist-4 DAs in Section~\ref{sec:4}.    
In Section~\ref{sec:5} we present numerical results for all DAs and
conclude in Section~\ref{sec:6} with a short summary and outlook. The appendices
contain a collection of relevant formulas, in particular the QCD
sum rules for the relevant twist-2, -3 and -4 matrix elements.

\section{General Framework and Twist-2 DAs}\label{sec:2}
\setcounter{equation}{0}

Light-cone meson DAs are defined in terms of matrix elements of non-local light-ray operators 
stretched along a certain light-like direction $z_\mu$, $z^2=0$, and
sandwiched between the vacuum and the meson state. We adopt the generic notation
\begin{equation}
    \phi_{t;M}(u),\ \psi_{t;M}(u),\ \ldots
\end{equation}
and 
\begin{equation}
    \Phi_{t;M}({\underline{\alpha}}),\ \Psi_{t;M}({\underline{\alpha}}),\ \ldots
\end{equation}
for two-particle and  three-particle DAs, respectively. The first subscript $t=2,3,4$ stands 
for the twist; the second one, $M=\pi,K,\ldots$, specifies the meson. 
{} For definiteness, we will write most of the expressions for $K$ mesons, 
i.e.\  $s\bar q$ bound states with $q =u,d$. The variable $u$ in the definition of two-particle DAs 
always refers to the momentum fraction carried by the quark, $u = u_s$; 
$\bar u \equiv 1-u = u_{\bar q}$ is the antiquark momentum fraction. 
The set of variables in the three-particle DAs,  
$\underline{\alpha} = \{\alpha_1,\alpha_2,\alpha_3\} = \{\alpha_{s},\alpha_{\bar q},\alpha_g\}$, corresponds 
to the momentum fractions carried by the quark, antiquark and gluon, respectively.    

To facilitate the light-cone expansion, it
is convenient to introduce a second light-like vector $p_\mu$ such that
\begin{equation}
p_\mu = P_\mu-\frac{1}{2}z_\mu \frac{m^2_M}{pz},
\label{smallp}
\end{equation}
where $P_\mu$ is the meson momentum, $P^2=m_M^2$. 
We also need the projector onto the directions orthogonal to $p$ and $z$,
\begin{equation}
       g^\perp_{\mu\nu} = g_{\mu\nu} -\frac{1}{pz}(p_\mu z_\nu+ p_\nu z_\mu),
\end{equation}
and use the notation
\begin{equation}
    a_z\equiv a_\mu z^\mu, \qquad  a_p\equiv a_\mu p^\mu, \qquad
    b_{\mu z} \equiv b_{\mu\nu} z^\nu, 
\qquad\mbox{\rm etc.}
\end{equation}
for arbitrary Lorentz vectors $a_\mu$ and tensors
$b_{\mu\nu}$. $a_\perp$ denotes the generic component of $a_\mu$ orthogonal to
$z$ and $p$.

We use the standard Bjorken--Drell
convention \cite{BD65} for the metric and the Dirac matrices; in particular,
$\gamma_{5} = i \gamma^{0} \gamma^{1} \gamma^{2} \gamma^{3}$,
and the Levi-Civita tensor $\epsilon_{\mu \nu \lambda \sigma}$
is defined as the totally antisymmetric tensor with $\epsilon_{0123} = 1$.
The covariant derivative is defined as
$D_{\mu} = \partial_{\mu} - igA_{\mu}$ and
the dual gluon-field-strength
tensor as $\widetilde{G}_{\mu\nu} =
\frac{1}{2}\epsilon_{\mu\nu \rho\sigma} G^{\rho\sigma}$.

The leading twist-2 DA of the $K$ meson is defined as%
\footnote{The leading-twist DA of a $\bar K$ meson
is given by 
$\phi_{2;\bar K}(u) = \phi_{2;K}(1-u)$.}
\begin{eqnarray}
\langle 0 | \bar q(z)[z,-z]\gamma_z\gamma_5 s(-z)|K(P)\rangle
& = & i f_K (pz) \int_0^1 du \, e^{i\xi pz} \, \phi_{2;K}(u,\mu^2)\,.
\label{eq:T2}
\end{eqnarray}
Here $[x,y]$ stands for the path-ordered gauge factor along the straight line
connecting the points $x$ and $y$:
\begin{equation}
[x,y] ={\rm P}\exp\left[ig\!\!\int_0^1\!\! dt\,(x-y)_\mu
  A_\mu(tx+(1-t)y)\right],
\label{Pexp}
\end{equation}
and $\mu$ is the renormalization (factorization) scale.
We also use the short-hand notation
\begin{equation}\label{eq:xi}
\xi = 2u-1.
\end{equation}
The decay constant $f_K$ is defined, as usual, as
\begin{equation}\label{eq:fP}
\langle 0|\bar q(0) \gamma_{\mu}\gamma_5
s(0)|K(P)\rangle = if_{K}P_\mu,
\end{equation}
with $f_K = 160\,$MeV \cite{PDG}.
The normalization follows from the requirement
that the local limit $z\to 0$ of (\ref{eq:T2}) reproduce
(\ref{eq:fP}), so that
\begin{equation}
\int_0^1 du\,\phi_{2;K}(u) = 1\,.
\end{equation}

A convenient tool to study DAs is provided by conformal expansion
\cite{BKM03}.\footnote{See Ref.~\cite{angi} for an alternative
  approach not based on conformal expansion.}
The underlying idea is similar to partial-wave decomposition in quantum mechanics and allows one to separate
transverse and longitudinal variables in the Bethe--Salpeter wave--function.  The
dependence on transverse coordinates is formulated as scale dependence
of the relevant operators and is governed by
renormalization-group equations, the dependence on the longitudinal
momentum fractions is described in terms of irreducible
representations of the corresponding symmetry group, the collinear
conformal group SL(2,$\mathbb R$). The conformal partial-wave expansion is
explicitly consistent with the equations of motion since the latter are
not renormalized. It thus makes maximum use of the symmetry
of the theory to simplify the dynamics. 

To construct the conformal expansion for an arbitrary multiparticle
distribution, one first has to decompose each constituent field into
components with fixed Lorentz-spin projection onto the
light-cone. Each such component has conformal spin
$$
j=\frac{1}{2}\, (l+s),
$$
where $l$ is the canonical dimension  and $s$ the (Lorentz-) spin
projection. In particular, $l=3/2$ for quarks and $l=2$ for gluons.
The  quark field is decomposed as $\psi_+ \equiv
\Lambda_+\psi$ and $\psi_-=
\Lambda_-\psi$ with spin projection operators $\Lambda_+ = \slash{p}\slash{z}/(2pz)$ and 
 $\Lambda_- = \slash{z}\slash{p}/(2pz)$, corresponding to
$s=+1/2$ and $s=-1/2$, respectively. For the gluon
field strength there are three possibilities:
$G_{z\perp}$ corresponds to $s=+1$,
$G_{p\perp}$ to $s=-1$, and both
$G_{\perp\perp}$ and $G_{zp}$ correspond to $s=0$.
Multiparticle states built of fields with definite Lorentz-spin
projection can be expanded in
irreducible  representations of SL(2,$\mathbb R$) 
with increasing conformal spin.
The explicit expression for the DA
of an $m$-particle state with the lowest possible conformal spin
 $j=j_1+\ldots+j_m$, the so-called asymptotic DA, is \cite{BF90}
\begin{equation}
\phi_{as}(\alpha_1,\alpha_2,\cdots,\alpha_m) =
\frac{\Gamma(2j_1+\cdots +2j_m)}{\Gamma(2j_1)\cdots \Gamma(2j_m)}\,
\alpha_1^{2j_1-1}\alpha_2^{2j_2-1}\ldots \alpha_m^{2j_m-1}.
\label{eq:asymptotic}
\end{equation}
Multiparticle irreducible representations with higher spin
$j+n,n=1,2,\ldots$,
are given by  polynomials of $m$ variables (with the constraint
$\sum_{k=1}^m \alpha_k=1$ ), which are orthogonal over
 the weight function (\ref{eq:asymptotic}).

In particular, for the leading-twist DA $\phi_{K;2}$ defined in
(\ref{eq:T2}), the expansion goes in Gegenbauer polynomials:
\begin{equation}\label{eq:confexp}
\phi_{K;2}(u,\mu^2) = 6 u (1-u) \left( 1 + \sum\limits_{n=1}^\infty
  a^K_{n}(\mu^2) C_{n}^{3/2}(2u-1)\right).
\end{equation}
To leading-logarithmic accuracy (LO), the (non-perturbative)
Gegenbauer moments 
$a_n$ renormalize multiplicatively with 
\begin{equation}
a^{\rm LO}_n(\mu^2) = L^{\gamma^{(0)}_n/(2\beta_0)}\, a_n(\mu_0^2),
\end{equation}
where $L\equiv \alpha_s(\mu^2)/\alpha_s(\mu_0^2)$, $\beta_0=(11N_c-2N_f)/3$, and
the anomalous dimensions $\gamma^{(0)}_n$ are given by
\begin{equation}
\gamma^{(0)}_n =  8C_F \left(\psi(n+2) + \gamma_E - \frac{3}{4} -
  \frac{1}{2(n+1)(n+2)} \right).
\end{equation}
The reason why leading-order renormalization respects the (anomalous)
conformal symmetry is that it is driven 
by tree-level counterterms that retain the symmetry properties of the Lagrangian.
More technically, the Callan--Symanzik equation that governs the renormalization-scale 
dependence coincides to this accuracy with the Ward identity for the dilatation operator, which is an 
element of the collinear conformal group \cite{BKM03}.

To next-to-leading order (NLO) accuracy, the scale dependence of the
Gegenbauer moments 
is more complicated and reads \cite{Mikhailov:1984ii,Mueller}
\begin{equation}
 a^{\rm NLO}_n(\mu^2) =  a_n(\mu_0^2) E_n^{\rm NLO} 
+\frac{\alpha_s(\mu^2)}{4\pi}\sum_{k=0}^{n-2} a_n(\mu_0^2)\,  E_k^{\rm NLO}\, d^{(1)}_{nk},  
\end{equation} 
where 
\begin{equation}
  E_n^{\rm NLO} =  L^{\gamma^{(0)}_n/(2\beta_0)}\left[1+ 
                  \frac{\gamma^{(1)}_n \beta_0 -\gamma_n^{(0)}\beta_1}{8\pi\beta_0^2}
                 \Big[\alpha_s(\mu^2)-\alpha_s(\mu_0^2)\Big]\right],
\end{equation}
$\gamma^{(1)}_n$ are the diagonal two-loop anomalous dimensions \cite{Floratos},
$\beta_1 = 102-(38/3)N_f$, 
and the mixing coefficients $d^{(1)}_{nk}$, $k\le n-2$, 
are given in closed form in Ref.~\cite{Mueller},
see also, for instance, Ref.~\cite{Bakulev:2004cu} for a recent compilation. 
For the lowest moments $n=0,1,2$ one needs
\begin{equation}
  \gamma_0^{(1)} =0\,, \qquad \gamma_1^{(1)} = 
\frac{23096}{243} - \frac{512}{81}\, N_f\,,
   \qquad  \gamma_2^{(1)} = 
 \frac{34450}{243}-\frac{830}{81}\, N_f
\end{equation}  
and
\begin{equation}
  d^{(1)}_{20} = \frac{7}{30}( 5 C_F - \beta_0)
              \frac{\gamma^{(0)}_2}{\gamma^{(0)}_2 -2 \beta_0}
                \left[1-L^{-1+\gamma^{(0)}_2/(2\beta_0)}\right].
\end{equation}

The odd Gegenbauer moments $a_{2n+1}$ are first order in  $SU(3)$-symmetry breaking 
for the kaon and vanish for the pion by virtue of G-parity. The numerical value of 
$a^K_1$ was the subject of significant controversy until recently. The existing estimates 
are all obtained using QCD sum rules. The first calculation of $a_1^K$ by Chernyak and 
Zhitnitsky yielded $a_1^K\approx 0.1$ \cite{CZZ82,CZreport}, but
unfortunately suffers from a sign mistake in the perturbative contribution \cite{BB03}. 
After the error is corrected, one finds that the two numerically 
leading contributions come with different sign and cancel to a large extent, so that the sum rule 
becomes unstable and numerically unreliable.
This problem was reanalysed in Refs.~\cite{KMM04,BL04,BZ05,BZ06} using
a different set of sum rules, where it was also checked that the
results are consistent with the equations of motion for the relevant 
operators \cite{BL04,BZ06}. The results are given in Table~\ref{tab:a1K}.  
\begin{table}[p]
\renewcommand{\arraystretch}{1.3}
\begin{center}
\begin{tabular}{l|l|l|l}\hline
Method       & $\mu=1$~GeV      & $\mu=2$~GeV    & Reference    \\ \hline
QCDSR,D      & $0.05 \pm 0.02$  &  $0.04\pm 0.02$  & \cite{KMM04} \\ 
QCDSR,ND;EOM & $0.10 \pm 0.12$  & $0.08\pm 0.09$   & \cite{BL04}  \\ 
QCDSR,D;EOM  & $0.06\pm 0.03$ &    $0.05\pm 0.02$  & \cite{BZ05,BZ06}  \\ \hline  
\end{tabular}
\end{center}
\caption[]{\sf The Gegenbauer moment $a_1^K(\mu^2)$ from QCD sum rules. The abbreviations stand for:
 QCDSR: QCD sum rules; D and ND:  diagonal and non-diagonal correlation function, respectively; 
 EOM: equations of motion. The  error estimates 
should be taken with some caution, as there is no systematic approach to estimate uncalculated 
higher-order terms in the OPE.}
\label{tab:a1K}
\vskip2cm
\renewcommand{\arraystretch}{1.0}
\renewcommand{\arraystretch}{1.3}
\begin{center}
\begin{tabular}{@{}l|l|l|l@{}} \hline
Method                      & $\mu=1$~GeV            & $\mu=2$~GeV      & Reference   
                                                                                      \\ \hline
CZ model                    & $ 0.56 $               & $0.38$        & \cite{Chernyak:1981zz,CZreport} 
                                                                                      \\ \hline 
QCDSR                       & $0.26^{+0.21}_{-0.09}$    &   $0.17^{+0.14}_{-0.06}$  & \cite{KMM04}  \\ 
QCDSR                       & $0.28\pm 0.08$            &   $0.19\pm 0.05$           & this paper    \\
QCDSR,NLC                   & $0.19\pm 0.06$         &   $0.13 \pm 0.04$  & \cite{Mikhailov:1991pt,Bakulev:1998pf,Bakulev:2001pa} 
                                                                                     \\ \hline
$F_{\pi\gamma\gamma^*}$,LCSR    &  $0.19 \pm 0.05$    & $0.12\pm 0.03 \,  (\mu=2.4)$ & \cite{Schmedding:1999ap} \\
$F_{\pi\gamma\gamma^*}$,LCSR    &  $0.32$  & $0.20\,(\mu=2.4)$ & \cite{Bakulev:2002uc} \\
$F_{\pi\gamma\gamma^*}$,LCSR,R  & $0.44$     & $0.30$   & \cite{Bakulev:2005cp}\\
$F_{\pi\gamma\gamma^*}$,LCSR,R  & $0.27$     & $0.18$   & \cite{Agaev:2005rc}
                                                                                      \\ \hline  
$F^{\rm em}_{\pi}$,LCSR        & $0.24\pm 0.14\pm 0.08$ &  $0.16\pm0.09\pm0.05$ & \cite{Braun:1999uj,Bijnens:2002mg}  \\
$F^{\rm em}_{\pi}$,LCSR,R      & $0.20\pm 0.03$         & $0.13\pm0.02$  & \cite{Agaev:2005gu}  
                                                                                      \\ \hline
$F_{B\to\pi\ell\nu}$,LCSR       & $0.19\pm 0.19$         & $0.13\pm0.13$              & \cite{BZ04}  
                                                                                      \\ \hline
LQCD, quenched,     & $0.381\pm 0.234^{+0.114}_{-0.062}$  &
                                                                                      $0.233\pm 0.143^{+0.088}_{-0.038}$    & $\mbox{\footnotesize UKQCD}$~\cite{DelDebbio:2002mq}  \\ 
 W/CW && $(\mu=2.67)$ & \\
LQCD, $N_f=2$, W/CW     &   $0.364\pm 0.126$   & $0.236\pm 0.082\, (\mu^2 = 5)$    & {\footnotesize QCDSF/UKQCD}~\cite{Gockeler:2005jz}  \\ \hline
\end{tabular}
\end{center}
\caption[]{\sf The Gegenbauer moment $a_2^\pi(\mu^2)$. 
 The CZ model involves $a_2^\pi =2/3$ at the low scale $\mu=500$~MeV; for the discussion of the 
 extrapolation to higher scales, see Ref.~\cite{Bakulev:2002uc}.  The abbreviations stand for: 
QCDSR: QCD sum rules; NLC: non-local condensates; LCSR: light-cone sum rules; R: renormalon model 
for twist-4 corrections; LQCD: lattice calculation; 
$N_f=2$: calculation using  $N_f=2$ dynamical quarks; W/WC: Wilson glue and 
non-perturbatively ${\mathcal O}(a)$ improved Clover--Wilson fermion action.
 }
\label{tab:a2pi}
\renewcommand{\arraystretch}{1.0}
\end{table}
As our best estimate, we take
\begin{equation}
a_1^K(1\,{\rm GeV}) = 0.06\pm 0.03.
\end{equation}

Calculations of the second Gegenbauer moment for the pion DA,
$a_2^\pi$, have attracted quite a bit of 
attention and have a long history. Three different approaches have
been used: direct calculations 
using QCD sum rules, pioneered by Chernyak and Zhitnitsky; analysis of
the experimental data on the 
pion electromagnetic and transition form factors and the $B$ weak decay
form factor, using light-cone sum rules; and lattice 
calculations. The summary of these results is presented in Table~\ref{tab:a2pi}; see also, 
for instance, Refs.~\cite{Bakulev:2002uc,Bakulev:2004cu} for another recent compilation. 

Our conclusion from Table~\ref{tab:a2pi} is rather pessimistic: $a_2^\pi$  can only be determined with
large errors, whatever approach is chosen. A fair quote is probably
\begin{equation}
 a_2^\pi(1\,{\rm GeV}) = 0.25\pm 0.15\,.
\label{a2pi}
\end{equation}

The $K$-meson DA has attracted comparatively less attention. The old estimate by Chernyak and Zhitnitsky, 
$\langle \xi^2\rangle_K/\langle \xi^2\rangle_\pi = 0.8\pm 0.02$ \cite{CZreport}, translates to 
\begin{equation}
                     a_2^K/a_2^\pi = 0.59\pm 0.04 \quad
                     \leftrightarrow \quad (a_2^K)_{\rm CZ}(1\,{\rm GeV}) = 0.33
\end{equation}
for the CZ model. A recent calculation, Ref.~\cite{KMM04}, including radiative corrections to the sum rules 
gives, however
\begin{equation}
      a_2^K/a_2^\pi \simeq 1\,, \qquad a_2^K(1\,{\rm GeV}) = 0.27^{+0.37}_{-0.12}.
\end{equation}
This result is consistent with the most recent lattice calculation, using $N_f=2$ dynamical fermions
\cite{Gockeler:2005jz}, which shows that $\langle\xi^2\rangle_\pi$
stays practically constant under a variation of the pion mass. For the
purpose of the present paper we have done an update of the QCD sum-rule calculation
\cite{KMM04}, using the corrected ${\mathcal O}(\alpha_s)$
quark-condensate contribution  given in Ref.~\cite{BZ05}, see
App.~\ref{app:C}, which yields
\begin{equation}
      a_2^K/a_2^\pi = 1.05\pm 0.15\,, \qquad a_2^K(1\,{\rm GeV}) = 0.30\pm 0.15\,.
\end{equation}
The difference with \cite{KMM04} is small and mainly due to the larger value of the strong coupling 
that we are using in this work. We conclude that the existing evidence points towards a very small 
$SU(3)$ violation in the second coefficient in the Gegenbauer expansion, so that we accept 
$a_2^K=a_2^\pi$ in the range given in Eq.~(\ref{a2pi}) as our final estimate.     

Estimates of yet higher-order Gegenbauer coefficients are rather uncertain. 
The light-cone sum-rule calculations of the transition form factor $F_{\pi\gamma\gamma^*}$ in 
Refs.~\cite{Schmedding:1999ap,Bakulev:2002uc,Bakulev:2005cp,Agaev:2005rc} suggest a negative
value for $a_4^\pi$, which is consistent with the result
$a_4^\pi(1\,{\rm GeV}) > -0.07$ obtained in Ref.~\cite{BZ04}.
However, this conclusion may be premature because of the omission of yet higher moments and 
absence of any convincing method to estimate systematic errors involved in the analysis. 
{}For this reason we adopt, in this paper, a model for the
leading-twist DA, which is given by the Gegenbauer expansion
(\ref{eq:confexp}) truncated after the second term.   

Last but not least
we have to specify the value of the strange-quark mass. In the last year several lattice calculations 
with dynamical fermions have been published; see
Refs.~\cite{Knechtli:2005ew,Gockeler:2006sh} for a summary and short review. In all these 
calculations the physical kaon mass is used as an input to fix the strange-quark mass. 
There is good agreement between data sets obtained using different non-perturbative renormalization procedures and, 
in fact, also with earlier quenched calculations. On the other hand,
the data still show considerable dependence 
on the lattice spacing, so that it is clear that simulations on finer
lattices are needed for a systematic continuum 
extrapolation.  In a different approach, the strange-quark mass can be extracted from
the $e^+e^-$ annihilation cross section and/or hadronic $\tau$-decay
data using QCD sum rules. 
These calculations have reached a certain degree of maturity and yield
results that are in reasonable agreement with lattice determinations; see  
Ref.~\cite{Narison:2005ny} for a recent summary and further references. In this paper we use
\begin{equation}
   \overline{m}_s(2\,{\rm GeV}) = (100\pm 20)\,\mbox{MeV},
\end{equation} 
which corresponds to $\overline{m}_s = (137\pm 27)\,\mbox{MeV}$ at 1~GeV. 

\section{Twist-3 Distributions}\label{sec:3}
\setcounter{equation}{0}

In this section we define all the twist-3 DAs of the kaon and derive
models for them to next-to-leading order in conformal expansion, which fulfil
the QCD equations of motion. We also work out the leading-order scale-dependence of the
corresponding hadronic parameters. Numerical values for the parameters
and the corresponding models are given in Section~\ref{sec:5}.

To twist-3 accuracy, there are two two-particle DAs defined as
\begin{eqnarray}
\langle 0 | \bar q(z) i\gamma_5 s(-z) | K(P)\rangle & = &
\frac{f_K m_K^2}{m_s+m_q}\, \int_0^1 du \, e^{i(2u-1) pz}\,
\phi^{p}_{3;K}(u),
\label{eq:2.11}\\
\langle 0 | \bar q(z) \sigma_{\alpha\beta}\gamma_5 s(-z) |
K(P)\rangle & = &-\frac{i}{3}\, \frac{f_K
  m_K^2}{m_s+m_q}  (P_\alpha z_\beta-
P_\beta z_\alpha) \int_0^1 du \, e^{i(2u-1) pz}\,\phi^{\sigma}_{3;K}(u).
\label{eq:2.12}
\end{eqnarray}
In addition, there is also one three-particle DA:
\begin{eqnarray}
\lefteqn{\langle 0 | \bar q(z) \sigma_{\mu\nu}\gamma_5
  gG_{\alpha\beta}(vz) s(-z)| K(P)\rangle\ =}
\hspace*{0.5cm}\nonumber\\
& = & i\,f_{3K} \left(p_\alpha p_\mu g_{\nu\beta}^\perp - p_\alpha p_\nu
  g_{\mu\beta}^\perp - (\alpha\leftrightarrow\beta)
 \right)\int {\cal D}\underline{\alpha} \, 
   e^{-ipz(\alpha_2
  -\alpha_1 + v\alpha_3)} {\Phi}_{3;K}(\alpha_1,\alpha_2,\alpha_3) +
  \dots,\nonumber\\[-15pt]\label{eq:3pT3}
\end{eqnarray}
where the integration measure is defined as
\begin{equation}
 \int {\cal D}\underline{\alpha}\ = \int_0^1 d\alpha_1\,d\alpha_2\,d\alpha_3\,\delta(1-\alpha_1-\alpha_2-\alpha_3)
\label{eq:measure}
\end{equation}
and the dots stand for Lorentz structures of twist 5 and higher.

To next-to-leading order in conformal spin, $\Phi_{3;K}$ is given by
\begin{equation}\label{eq:turk}
{\Phi}_{3;K}(\underline{\alpha}) = 360 \alpha_1\alpha_2\alpha_3^2
\left\{ 1 + \lambda_{3K} (\alpha_1-\alpha_2) + 
\omega_{3K}\, \frac{1}{2}\left( 7\alpha_3-3\right)\right\}.
\end{equation}
The three parameters
$f_{3K}$, $\lambda_{3K}$, and $\omega_{3K}$ can be defined in terms of
matrix elements of local twist-3 operators as follows:
\begin{eqnarray}
\langle 0 | \bar q \sigma_{z\xi}\gamma_5 gG_{z\xi}
s|K\rangle
& = & 2 i f_{3K} (pz)^2,
\nonumber\\
\langle 0 | \bar q \sigma_{z\xi} \gamma_5 [iD_z,gG_{z\xi}] s
- \frac{3}{7}\, i\partial_z \bar q \sigma_{z\xi} \gamma_5
gG_{z\xi} s | K\rangle
& = & 2i f_{3K} (pz)^3
\,\frac{3}{28}\, \omega_{K3},
\nonumber\\
\langle 0 | \bar q i \!\stackrel{\leftarrow}{D}_z\!
\sigma_{z\xi} \gamma_5 gG_{z\xi} s - \bar q
\sigma_{z\xi} \gamma_5 gG_{z\xi} i\! \stackrel{\rightarrow}{D}_z\! s
  | K\rangle & = & 2i f_{3K} (pz)^3
\,\frac{1}{14}\, \lambda_{K3}.
\label{eq:T3ops}
\end{eqnarray}
Numerical values for these parameters can be obtained from QCD sum
rules and will be discussed in Section~\ref{sec:5}.

The operators in (\ref{eq:T3ops}) renormalize multiplicatively in the chiral 
limit, with one-loop anomalous dimensions~\cite{adim3}  
\begin{eqnarray}\label{eq:1-loop}
\gamma_{3;f}^{(0)} &=& 2 C_A + \frac{14}{3}\,C_F =
\frac{110}{9},\nonumber\\ 
\gamma_{3;\omega}^{(0)} &=& \frac{20}{3}\,C_A +
\frac{7}{3}\,C_F = \frac{208}{9}, \nonumber\\
\gamma_{3;\lambda}^{(0)} &=& \frac{5}{3}C_A + \frac{47}{6}C_F =\frac{139}{9}\,.
\end{eqnarray}
{}For a massive strange quark, the operators in (\ref{eq:T3ops}) mix with
twist-2 ones. Using the light-ray-operator technique of
Ref.~\cite{string}, this mixing can be expressed in compact form as 
\begin{eqnarray}
   O_3(z,vz,0)^{\mu^2} &=&  O_3(z,vz,0)^{\mu_0^2} - im_s 
\,\frac{C_F\alpha_s}{2\pi}\,\ln\,\frac{\mu^2}{\mu_0^2}
 \, \frac{1}{v} \int_0^1\!\! dt\,\Big[O_2(z,vz) - 2 t O_2(z,tvz)\Big],\hspace*{10pt}
  \label{eq:OPmix} 
\end{eqnarray}
where 
\begin{equation}
  O_3(z,vz,0)^{\mu^2} = 
[\bar q(z)\sigma_{z\nu}\gamma_5 g G_{z\nu}(vz) s(0)]^{\mu^2}
\end{equation} 
and 
\begin{equation}
    O_2(az,bz)^{\mu^2} = [\bar q(az)\gamma_z \gamma_5 s(bz)]^{\mu^2};
\end{equation} 
$\mu^2$ stands for the normalization point. Sandwiching
(\ref{eq:OPmix}) between the $K$ state and the vacuum, and expanding in powers of $pz$, one can easily derive
the mixing for local operators with an arbitrary number of
derivatives. We find that $f_{3K}$ mixes with $f_K m_s$ and with $f_K
m_s a_1^K$, whereas $\lambda_{3K}$ and $\omega_{3K}$ mix in addition
with $f_K m_s a_2^K$. The corresponding LO renormalization-group-improved expressions read
\begin{eqnarray}
f_{3K}(\mu^2) & = & L^{55/(9\beta_0)} f_{3K}(\mu_0^2) +
\frac{2}{19}\left(L^{4/(\beta_0)} - L^{55/(9\beta_0)}\right)
f_K m_s(\mu_0^2)
\nonumber\\
&&{} + \frac{6}{65}\left(L^{55/(9\beta_0)} -
L^{68/(9\beta_0)}\right) f_K [m_s a_1^K](\mu_0^2),
\nonumber\\
{}[f_{3K}\omega_{3K}](\mu^2) & = & L^{104/(9\beta_0)} 
[f_{3K}\omega_{3K}](\mu_0^2) +
\frac{1}{170}\left(L^{4/(\beta_0)} - L^{104/(9\beta_0)}\right)
f_K m_s(\mu_0^2)
\nonumber\\
&&{} + \frac{1}{10}\left(L^{68/(9\beta_0)}-L^{104/(9\beta_0)}
\right) f_K [m_s a_1^K](\mu_0^2)
\nonumber\\
&&{}+ \frac{2}{15}\left(L^{86/(9\beta_0)}-L^{104/(9\beta_0)} 
\right) f_K [m_s a_2^K](\mu_0^2), 
\nonumber\\
{}[f_{3K}\lambda_{3K}](\mu^2) & = & L^{139/(18\beta_0)} 
[f_{3K}\lambda_{3K}](\mu_0^2) -
\frac{14}{67}\left(L^{4/(\beta_0)} - L^{139/(18\beta_0)}\right)
f_K m_s(\mu_0^2)
\nonumber\\
&&{} + \frac{14}{5}\left(L^{68/(9\beta_0)}-L^{139/(18\beta_0)}
\right) f_K [m_s a_1^K](\mu_0^2)
\nonumber\\
&&{}- \frac{4}{11}\left(L^{86/(9\beta_0)}-L^{139/(18\beta_0)} 
\right) f_K [m_s a_2^K](\mu_0^2), \label{eq:evolution3}
\end{eqnarray}
where $L$ is the leading-log scaling factor: $L=\alpha_s(\mu^2)/\alpha_s(\mu_0^2)$. 

The two-particle twist-3 DAs (\ref{eq:2.11}) and  (\ref{eq:2.12}) are not independent,
but related to the three-particle DA $\Phi_{3;K}$ by EOM \cite{BF90,PB98}.
The EOM relations contain terms that depend on quark masses and can conveniently be 
expressed in terms of two dimensionless parameters $\rho^K_\pm$:
\begin{equation}
    \rho_{+}^K =\frac{(m_s+m_q)^2}{m^2_K},\qquad \rho_{-}^K =
\frac{m_s^2-m_q^2}{m_K^2}\,;
\end{equation}
numerically $ \rho_{+}^K \simeq  \rho_{-}^K$. The rationale for introducing two parameters is that 
$\rho_{-}^K$ changes sign when switching from $K$ mesons to $\bar K$ mesons, 
i.e.\ $ \rho_{+}^{\bar K} =  \rho_{+}^K$, but $ \rho_{-}^{\bar K} = -\rho_{-}^K$. 
In the analysis of twist-3 DAs given in Ref.~\cite{PB98}, only terms in $\rho_+^K$ have been included. 
Here we complete these studies by taking into account also the terms in $\rho_-^K$.

{}From the non-local operator identities (\ref{eq:oprel3}) and (\ref{eq:oprel4}), 
one obtains the following relations for moments of the DAs, dropping
the index $K$:
\begin{eqnarray}
M_n^{\phi_3^p} & = & \delta_{n0} + \frac{n-1}{n+1}\, M^{\phi_3^p}_{n-2} + 2 (n-1)
M^{{\varphi}_3^{(1)}}_{n-2} + \frac{2(n-1)(n-2)}{n+1}\, M^{{\varphi}_3^{(2)}}_{n-3}
\nonumber\\
&&{}- \rho_+ \,\frac{n-1}{n+1}\, M_{n-2}^{\phi_2} + \rho_- M_{n-1}^{\phi_2}\,,
\nonumber\\
M_n^{\phi_3^\sigma} & = & \delta_{n0} + \frac{n-1}{n+3}\, M^{\phi_3^\sigma}_{n-2} + \frac{6
  (n-1)}{n+3} \, M^{{\varphi}_3^{(1)}}_{n-2} + \frac{6n}{n+3}\, M^{{\varphi}_3^{(2)}}_{n-1}
\nonumber\\
&&{}- \rho_+ \,\frac{3}{n+3}\, M_n^{\phi_2} + \rho_-\,\frac{3}{n+3}\,M_{n-1}^{\phi_2}\,,
\label{eq:momrel3}
\end{eqnarray}
where we use the notation 
$$M_n^{\phi} = \int_0^1 du\, (2u-1)^n \phi(u)$$ 
and introduce the auxiliary functions
\begin{eqnarray}
\varphi_3^{(1)}(u) &=& \int_0^u d\alpha_1 \int_0^{\bar u} d\alpha_2
\,\frac{2}{1-\alpha_1-\alpha_2} \,{\Phi}_3(\underline{\alpha})\,,\\
\varphi_3^{(2)}(u) &=& \int_0^u d\alpha_1 \int_0^{\bar u} d\alpha_2
\,\frac{2}{(1-\alpha_1-\alpha_2)^2} \,(\alpha_1-\alpha_2-(2u-1))\,
{\Phi}_3(\underline{\alpha})\,.
\end{eqnarray}
 The normalization is chosen in such a way that
\begin{equation} 
M_0^{\phi_3^p} = \int\limits_0^1 du\, \phi_3^p(u) = 1,\qquad M_0^{\phi_3^\sigma} = \int\limits_0^1 du
\,\phi_3^\sigma(u) = 1 - \rho_+.
\end{equation}
Except for the new terms in $\rho_-$, these moment relations
agree with those obtained in Refs.~\cite{BF90,PB98}. 

The relations (\ref{eq:momrel3}) can be solved
exactly: separating the contributions of quark--an\-ti\-quark--glu\-on operators and the terms in $\rho_\pm$, 
\begin{eqnarray}
\phi_3^{p}(u) &=& 1+ \phi^{p}_{3,g}(u) + \rho_+ \,\phi^{p}_{3,+}(u) + \rho_- \,\phi^{p}_{3,-}(u)\,,
\nonumber\\
\phi_3^{\sigma}(u) &=& 6u\bar u+ \phi^{\sigma}_{3,g}(u) + \rho_+
\,\phi^{\sigma}_{3,+}(u) + \rho_- \,\phi^{\sigma}_{3,-}(u)\,,
\end{eqnarray}
we obtain the integral representations (cf.\ Ref.~\cite{SCET})
\begin{eqnarray}
\phi^p_{3,g}(u) & = &\frac14\int_0^u \frac{dv}{\bar v}\left[
(2v-1) \,(\varphi_3^{(1)})^{''} - 2 \,(\varphi_3^{(1)})^{'}(v) + (\varphi_3^{(2)})^{''}(v)\right]
\nonumber\\
&&{} 
- \frac14\int_u^1\frac{dv}{v}\left[
(2v-1) \,(\varphi_3^{(1)})^{''}(v) - 2 \,(\varphi_3^{(1)})^{'}(v) + (\varphi_3^{(2)})^{''}(v)\right],
\label{eq:phig}
\end{eqnarray}
\begin{eqnarray}
\phi^p_{3,+}(u) & = & \frac14\int_0^u \frac{dv}{\bar v}\,\phi_2'(v) 
- \frac14\int_u^1 \frac{dv}{v}\,\phi_2'(v)\,,
\label{eq:phiplus}\\
\phi^p_{3,-}(u) & = & \frac14\int_0^u \frac{dv}{\bar v}\,\left[2 \phi_2(v) - \phi_2'(v)\right] - 
\frac14\int_u^1 \frac{dv}{v}\,\left[2 \phi_2(v) +\phi'_2(v)\right],
\label{eq:phiminus}
\end{eqnarray}
where primes denote 
the derivatives in $v$: $\phi'(v) = (d/dv)\phi(v)$ etc.

For the second twist-3 DA the solutions of the moment relations read,
in the same notation:
\begin{eqnarray}
\phi^\sigma_{3,g}(u) & = &-\frac{3}{2}\,u\bar u \left\{ \int_0^u
dv\left( \frac{1}{\bar v^2} + \frac{2}{\bar v}\right) \left(
(\varphi_3^{(1)})^{'}(v) + (2v-1)\,(\varphi_3^{(2)})^{'}(v)
\right)\right.\\
&&\left. \hspace*{0.8cm}{}- \int_u^1
dv\left( \frac{1}{ v^2} + \frac{2}{ v}\right) \left(
(\varphi_3^{(1)})^{'}(v) + (2v-1)\,(\varphi_3^{(2)})^{'}(v)
\right)\right\},\\
\phi^\sigma_{3,+}(u) & = & -\frac{3}{2}\,u\bar u
\left(\int_0^u dv\,\frac{1}{\bar
  v^2}\,\phi_2(v) + \int_u^1
dv\,\frac{1}{v^2}\,\phi_2(v)\right),
\nonumber\\
\phi^\sigma_{3,-}(u) & = & \frac{3}{2}\,u \bar u \left\{
\int_0^u dv\,\left(\frac{1}{\bar
  v^2} + \frac{2}{\bar v}\right) \phi_2(v) - 
\int_u^1 dv\,\left(\frac{1}{
  v^2} + \frac{2}{v}\right) \phi_2(v)\right\}. 
\label{eq:phisigma}
\end{eqnarray}

We stress that the relations (\ref{eq:phig}) to (\ref{eq:phisigma}) are valid in full QCD and 
involve no approximation whatsoever.
One consequence of these relations is that quark-mass corrections 
to $\phi_3^{p,\sigma}$ contain logarithmic end-point singularities. 
In particular for the asymptotic leading-twist DA $\phi_{2;K}(u) =6u(1-u)$ we obtain
\begin{equation}
\label{eq:disaster}
\phi^p_{3;K}(u)_{{\mbox{$|$}}_{\mbox{\scriptsize no gluons, asymptotic $\phi_{2;K}$}}} 
= 1 + \rho_+^K \,\frac{3}{2}\,(
2 + \ln u \bar u) + \rho_-^K \,\frac{3}{2} \left(1-2 u +
\ln\,\frac{u}{\bar u}\right).
\end{equation}
To NLO in conformal spin we obtain, using the truncated conformal expansions (\ref{eq:confexp})
for $\phi_{2;K}$ and (\ref{eq:turk}) for $\Phi_{3;K}$:
\begin{eqnarray}
\lefteqn{\phi^p_{3;K}(u) = 1 + 3\rho^K_+ (1+6 a^K_2) - 9 \rho^K_- a^K_1 +
C_1^{1/2}(2u-1) \left[\frac{27}{2}\,\rho^K_+ a^K_1 - \rho^K_- \left(
\frac{3}{2} + 27 a^K_2\right)\right]}
\nonumber\\
&& + C_2^{1/2}(2u-1) \left( 30 \eta_{3K} + 15 \rho^K_+ a^K_2 - 3
\rho^K_- a^K_1\right) + C_3^{1/2}(2u-1)\left( 10 \eta_{3K} \lambda_{3K}
- \frac{9}{2}\,\rho^K_- a^K_2\right)
\nonumber\\
&&{} - 3\eta_{3K} \omega_{3K} C_4^{1/2}(2u-1) + \frac{3}{2}\,(\rho^K_+
+ \rho^K_-) (1-3a^K_1+6a^K_2)\ln u
\nonumber\\
&&{} + \frac{3}{2}\,(\rho^K_+-\rho^K_-)(1+3 a^K_1 + 6  a^K_2) \ln \bar u,\label{eq:phip}\\ 
\lefteqn{\phi^\sigma_{3;K}(u) =  6u\bar u \left[ 1 + \frac{3}{2}\,\rho^K_+ + 15
\rho^K_+ a^K_2 - \frac{15}{2}\,\rho^K_- a^K_1 + \left( 3 \rho^K_+ a^K_1 -
\frac{15}{2}\,\rho^K_- a^K_2\right) C_1^{3/2}(2u-1)\right.}
\nonumber\\
&&{}\left. + 
    \left(5\eta_{3K} -\frac{1}{2}\,\eta_{3K}\omega_{3K}  +
    \frac{3}{2}\,\rho^K_+ a^K_2 \right)
    C_2^{3/2}(2u-1) + \eta_{3K} \lambda_{3K}
    C_3^{3/2}(2u-1)\right]
\nonumber\\
&&{} +  9 u \bar u (\rho^K_++\rho^K_-) (1-3 a^K_1+6 a^K_2) \ln u + 
9 u \bar u (\rho^K_+-\rho^K_-) (1+3 a^K_1+6 a^K_2) \ln \bar u\,,
\label{eq:3part}
\end{eqnarray}
where, to simplify notations, we have introduced the parameter 
\begin{equation}
\eta_{3K} =  \frac{f_{3K}}{f_K}\,\frac{m_q+m_s}{m_K^2}.
\end{equation}
These expressions are our final results for the two-particle twist-3
DAs and supersede those given in Refs.~\cite{BF90,PB98}. The terms
multiplying $\ln u$ and $\ln\bar u$ are the first three terms in the
conformal expansion of $\phi'_{2;K}(0)$ and $\phi'_{2;K}(1)$, respectively.
Numerical values for the hadronic parameters are given in Table~\ref{tab:num}. 
The leading-order scale-dependence follows from (\ref{eq:evolution3}) and the
scale dependence of the quark masses in $\rho^K_\pm$ and  $\eta_{3K}$.

We note in passing that the EOM relations
\begin{eqnarray*}
\frac{u}{2}\left\{\phi_{3;\pi}^{p}(u)+\frac{1}{6}\,(\phi_{3;\pi}^\sigma(u))'\right\}_{\mbox{\tiny
    no gluons}} &=& \left.\frac{1}{6}\,\phi_{3;\pi}^\sigma(u)\right|_{\mbox{\tiny
    no gluons}}\,,\\
\frac{1-u}{2}\left\{\phi_{3;\pi}^{p}(u)-\frac{1}{6}\,(\phi_{3;\pi}^\sigma(u))'\right\}_{\mbox{\tiny
    no gluons}} &=& \left.\frac{1}{6}\,\phi_{3;\pi}^\sigma(u)\right|_{\mbox{\tiny
    no gluons}}\,,
\end{eqnarray*}
are no longer fulfilled for $\phi_{3;K}^{p,\sigma}$, but violated by mass corrections in $\rho_\pm^K$.

\section{Twist-4 Distributions}\label{sec:4}
\setcounter{equation}{0}

In this section we derive models for the two- and three-particle
twist-4 DAs to NLO in the conformal expansion.
There are four $K$-meson three-particle DAs of twist 4, defined as~\cite{BF90,PB98}%
\footnote{
   In  the notation of  Ref.~\cite{PB98},
   $\Phi_{4;K} =m^2_K {\mathcal A}_\parallel$,
   $\Psi_{4;K}  = m^2_K  {\mathcal A}_\perp$,  
   $\widetilde\Phi_{4;K}  = m^2_K  {\mathcal V}_\parallel$ and
   $\widetilde\Psi_{4;K}  = m^2_K  {\mathcal V}_\perp$.}  
\begin{eqnarray}
\lefteqn{\langle 0 | \bar q(z)\gamma_\mu\gamma_5
gG_{\alpha\beta}(vz)s(-z)|K(P)\rangle\ =}
\hspace*{0.5cm}\nonumber\\
& = & p_\mu (p_\alpha z_\beta - p_\beta z_\alpha)\, \frac{1}{pz}\, f_{K} 
\Phi_{4;K}(v,pz) + (p_\beta g_{\alpha\mu}^\perp -
p_\alpha g_{\beta\mu}^\perp) f_{K} \Psi_{4;K}(v,pz) + \dots,
\label{4.15}\\
\lefteqn{\langle 0 | \bar q(z)\gamma_\mu i
g\widetilde{G}_{\alpha\beta}(vz)s(-z)| K(P)\rangle\ =}
\hspace*{0.5cm}\nonumber\\
& = & p_\mu (p_\alpha z_\beta - p_\beta z_\alpha)\, \frac{1}{pz}\, f_K
\widetilde\Phi_{4;K}(v,pz) + (p_\beta g_{\alpha\mu}^\perp -
p_\alpha g_{\beta\mu}^\perp) f_{K} \widetilde\Psi_{4;K}(v,pz) + \dots,
\label{4.16}
\end{eqnarray}
with the short-hand notation
$${\cal F}(v,pz) = \int{\cal D}\underline{\alpha}\,
e^{-ipz(\alpha_2-\alpha_1+v \alpha_3)} {\cal F}(\underline{\alpha}).$$
The integration measure ${\cal D}\underline{\alpha}$ is defined in (\ref{eq:measure}), 
and the dots denote terms of twist 5 and higher.
{}For massless quarks and, more generally, for two equal-mass quarks, G-parity implies that
 the DAs $\Phi$ and $\Psi$ are antisymmetric under the interchange of the quark 
momenta, $\alpha_1\leftrightarrow \alpha_2$, whereas $\widetilde\Phi$ and 
$\widetilde\Psi$ are symmetric \cite{BF90,PB98}.
Note that unlike twist-2 and twist-3 DAs, which are dimensionless, 
the twist-4 DAs have mass dimension 2 (GeV$^2$). The corresponding contributions to hard exclusive 
processes are suppressed by two powers of the hard scale with respect to leading twist. 

The distribution amplitudes $\Phi_{4;K}$ and  $\widetilde \Phi_{4;K}$ correspond to the light-cone projection
$\gamma_z G_{z p},$ which picks up the $s=+1/2$ components of both
quark and antiquark field and the $s=0$ component of
 the gluon field. The conformal expansion reads:
\begin{eqnarray}
\Phi_{4;K}(\underline{\alpha}) & = & 120
\alpha_1\alpha_2\alpha_3 [ \phi_0^K + \phi_1^K(\alpha_1-\alpha_2) + \phi_2^K (3\alpha_3-1) +
\dots],\nonumber\\
\widetilde\Phi_{4;K}(\underline{\alpha}) & = & 120
\alpha_1\alpha_2\alpha_3 [ \widetilde\phi_0^K +
  \widetilde\phi_1^K(\alpha_1-\alpha_2) + 
\widetilde\phi_2^K (3\alpha_3-1) +
\dots].
\label{eq:phi}
\end{eqnarray}
G-parity implies that, for the $\pi$ meson, $\phi_0^\pi = \phi_2^\pi = \widetilde\phi_1^\pi = 0$, 
whereas  $\phi_0^K$, $\phi_2^K$ and $\widetilde\phi_1^K$ are ${\mathcal O}(m_s-m_q)$.

In turn, the DAs $\Psi_{4;K}$ and  $\widetilde \Psi_{4;K}$ correspond to the light-cone projection
$\gamma_\perp G_{z \perp}$, which is a mixture of different quark-spin states with $s_q=+1/2, s_{\bar q}=-1/2$
and  $s_q=-1/2, s_{\bar q}=+1/2$, respectively. In both cases $s=+1$ for the gluon.  
We separate the different quark-spin
projections by introducing the auxiliary amplitudes ${\Psi}^{\uparrow\downarrow}$ and 
${\Psi}^{\downarrow\uparrow}$, defined as
\begin{eqnarray}
\langle 0 | \bar q(z) ig\widetilde{G}_{\alpha\beta}(vz) \gamma_z
\gamma_\mu\gamma_p s(-z) | K(P)\rangle & = & f_K
\left( p_\beta g^\perp_{\alpha\mu} - p_\alpha
  g^\perp_{\beta\mu} \right)
{\Psi}^{\uparrow\downarrow}(v,pz),
\nonumber\\
\langle 0 | \bar q(z) ig\widetilde{G}_{\alpha\beta}(vz) \gamma_p
\gamma_\mu\gamma_z s(-z) | K(P)\rangle & = & f_K
\left( p_\beta g^\perp_{\alpha\mu} - p_\alpha
  g^\perp_{\beta\mu} \right)
{\Psi}^{\downarrow\uparrow}(v,pz).
\end{eqnarray}
The original distributions  $\Psi_{4;K}$ and  $\widetilde \Psi_{4;K}$ are  given by
\begin{eqnarray}
{\widetilde\Psi}(\underline{\alpha}) & = & -\frac{1}{2}\Big[ {\Psi}^{
\uparrow\downarrow}(\underline{\alpha}) + {\Psi}^{\downarrow\uparrow}(\underline{\alpha})\Big],\quad
{\Psi}(\underline{\alpha}) = 
\frac{1}{2}\Big[ {\Psi}^{
\uparrow\downarrow}(\underline{\alpha}) - {\Psi}^{\downarrow\uparrow}(\underline{\alpha})\Big].
\label{eq:original}
\end{eqnarray}
${\Psi}^{\uparrow\downarrow}$ and ${\Psi}^{\downarrow\uparrow}$
have a regular expansion in terms of conformal  polynomials, to wit:
\begin{eqnarray}
{\Psi}^{\uparrow\downarrow}(\underline{\alpha}) & = &
60\alpha_2\alpha_3^2 \left[ \psi_{0}^{\uparrow\downarrow} + 
\psi_{1}^{\uparrow\downarrow}(\alpha_3-3\alpha_1) +
\psi_{2}^{\uparrow\downarrow}\left(\alpha_3-\frac{3}{2}\alpha_2\right)\right],
\nonumber\\
{\Psi}^{\downarrow\uparrow}(\underline{\alpha}) & = &
60\alpha_1\alpha_3^2 \left[ \psi_{0}^{\downarrow\uparrow} +
\psi_{1}^{\downarrow\uparrow}(\alpha_3-3\alpha_2) +
  \psi_{2}^{\downarrow\uparrow}\left(\alpha_3-\frac{3}{2}\alpha_1\right)\right].
\end{eqnarray}
For the $\pi$ meson, thanks to G-parity,
\begin{equation}
{\Psi}^{\uparrow\downarrow}_{4;\pi}(\alpha_1,\alpha_2)={\Psi}^{\downarrow\uparrow}_{4;\pi}(\alpha_2,\alpha_1),
\end{equation}
so that $\psi_i^{\uparrow\downarrow}\equiv \psi_i^{\downarrow\uparrow}$.%
\footnote{This implies, in particular, that only one of the DAs $\Psi$ and $\widetilde\Psi$
is dynamically independent.}  For
$K$, we write
\begin{equation}
\psi_i^{\uparrow\downarrow} = \psi^K_i + \theta^K_i,\qquad 
\psi_i^{\downarrow\uparrow} = \psi^K_i - \theta^K_i,
\end{equation}
where the $\theta_i$ correspond to  $SU(3)$-breaking corrections that 
also violate G-parity.
{}From (\ref{eq:original}), the following representations can readily be derived:
\begin{eqnarray}
{\widetilde\Psi}_{4;K}(\underline{\alpha}) & = &
 -30 \alpha_3^2\bigg\{ \psi^K_{0}(1-\alpha_3)
                    +\psi^K_{1}\Big[\alpha_3(1-\alpha_3)-6\alpha_1\alpha_2\Big]
\label{eq:tildepsi}\\
& & 
                    +\psi^K_{2}\Big[\alpha_3(1-\alpha_3)-\frac{3}{2}(\alpha_1^2
                               +\alpha_2^2)\Big]-
		    (\alpha_1-\alpha_2)\left[ \theta^K_0 + \alpha_3
		      \theta^K_1 + \frac{1}{2}\,(5 \alpha_3-3)
		      \theta^K_2\right]\bigg\},
\nonumber\\
 {\Psi}_{4;K}(\underline{\alpha}) & = &
 30 \alpha_3^2\bigg\{ \theta^K_{0}(1-\alpha_3)
                    +\theta^K_{1}\Big[\alpha_3(1-\alpha_3)-6\alpha_1\alpha_2\Big]
\label{eq:psi}\\
& &
                    +\theta^K_{2}\Big[\alpha_3(1-\alpha_3)-\frac{3}{2}(\alpha_1^2
                               +\alpha_2^2)\Big]-
		    (\alpha_1-\alpha_2)\left[ \psi^K_0 + \alpha_3
		      \psi^K_1 + \frac{1}{2}\,(5 \alpha_3-3)
		      \psi^K_2\right]\bigg\}.
\nonumber
\end{eqnarray}
In addition, we introduce one more three--particle DA $\Xi_{4}(\underline{\alpha})$ \cite{BGG04}:
\begin{equation}
\label{Xi} 
\left\langle 0\left \vert \bar{q}(z)
\gamma_{\mu}  \gamma_{5} gD^{\alpha}G_{\alpha\beta}(vz)s(-z)
\right\vert K^+(p)\right\rangle
=
if_{K}p_{\mu}p_{\beta}
\int \!{\cal D}\underline{\alpha}\,\,
{\rm e}^{-ipz(\alpha_2-\alpha_1+v\alpha_3)}\,\Xi_{4;K}(\underline{\alpha}).
\end{equation}
The Lorentz structure $p_{\mu}p_{\beta}$ is the only one relevant
at twist 4. Because of the EOM,
$D^{\alpha}G^A_{\alpha\beta} = -g\sum_q \bar q t^A \gamma_\beta q$,
where the summation goes over all light flavors, $\Xi_{4;K}(\underline{\alpha})$
can be viewed as describing  either a quark--antiquark--gluon or a
specific four--quark component of the pion, with the
quark--antiquark pair in a color--octet state and at the same
space--time point. 
The conformal expansion of $\Xi_{4;K}(\underline{\alpha})$ starts with $J=4$ and reads
\begin{equation}
  \Xi_{4;K}(\underline{\alpha}) = 840 \alpha_1\alpha_2\alpha_3^3\Big[\Xi^K_0 +\ldots\Big],
\label{Xi1}
\end{equation}
where $\Xi^K_0$  has mass dimension 2. The dots stand for terms with  
higher conformal spin $J=5,6,\ldots$, which are beyond our accuracy.
This DA was not considered in Refs.~\cite{BF90,PB98} because  $\Xi^K_0
={\cal O}(m_s-m_q)$ and
vanishes for mesons built of quark and antiquark with equal mass.

The expressions in Eqs.~(\ref{eq:phi}), (\ref{eq:tildepsi}),
(\ref{eq:psi}), (\ref{Xi}) represent the most general
parametrization of the twist-4 DAs to  NLO in the conformal-spin expansion and involve 
13 non-per\-tur\-ba\-tive parameters. Not all of them are independent,
though. In the following, we shall establish their mutual relations
and also  express the expansion coefficients in terms of matrix elements of local operators.

The asymptotic three-particle DAs correspond to contributions of the lowest conformal spin 
$J = j_s+j_{\bar q}+ j_{g} = 3$. The parameters 
$\phi_0^K$, $\widetilde\phi_0^K$, $\psi_0^K$ and $\theta_0^K$
describing these DAs can be
expressed in terms of local matrix elements as 
\begin{eqnarray}
\langle 0 | \bar q\gamma_\alpha
g\widetilde{G}_{\mu\alpha}s|K(P)\rangle &=& i P_\mu f_{K}\delta^2_K,
\nonumber\\
\langle 0 | \bar q\gamma_\alpha\gamma_5 igG_{\mu\alpha}s|K(P)\rangle &=& i P_\mu f_{K} m_K^2 \kappa_{4K}.
\label{eq:def33}
\end{eqnarray}
These are the only two local twist-4 operators of dimension 5. Note
that the second matrix element vanishes for equal-mass quarks, because
of G-parity. It also
vanishes  in the chiral limit $m_q,m_s\to 0$ because of the factor $m_K^2$.
Moreover, in this limit $\kappa_{4K}$ can be calculated exactly to leading order 
in $m_s$ \cite{BL04}:
\begin{equation}
    \kappa_{4K} = -\frac18 + {\mathcal O}(m_s);
\label{eq:kappa4K}
\end{equation}
numerical estimates of the corrections can be obtained from QCD sum rules and will be discussed below. 

Taking the local limit of  Eqs.~(\ref{4.15}), (\ref{4.16}), one obtains
\begin{equation}
\psi^K_{0} =  \widetilde\phi^K_{0} = -\frac{1}{3}\,\delta^2_K,\qquad
\phi^K_0 = -\theta^K_0 = \frac{1}{3}\,m_K^2 \kappa_{4K}.
\label{eq:theta0}
\end{equation}
What about the scale-dependence of these parameters?
Like $f_{3K}$, $\delta_K^2$
renormalizes multiplicatively for massless quarks, but mixes with operators
of lower twist for $m_s\neq 0$. At the operator level, neglecting 
${\mathcal O}(m_s^2)$ corrections, the mixing is
given by
\begin{equation}\label{T4mix}
(\bar q \gamma_\alpha g \widetilde{G}_{\mu\alpha}s)^{\mu^2} = (\bar q
  \gamma_\alpha g 
\widetilde{G}_{\mu\alpha}s)^{\mu_0^2}
  \left(1-\frac{8}{9}\,\frac{\alpha_s}{\pi}\,\ln\,\frac{\mu^2}{\mu_0^2}
  \right) -
  \frac{1}{9}\,\frac{\alpha_s}{\pi}\,\ln\,\frac{\mu^2}{\mu_0^2}\, m_s
  \left[\partial_\mu (\bar q i \gamma_5 s)\right]^{\mu_0^2}.
\end{equation}
Taking matrix elements and resumming the logarithms, we find
\begin{equation}
[\delta_K^2](\mu^2) = L^{32/(9\beta_0)}[\delta_K^2](\mu_0^2) +
\frac{1}{8}\,\left(1-L^{32/(9\beta_0)}\right) m_K^2,
\end{equation}
with, as before, $L = \alpha_s(\mu^2)/\alpha_s(\mu_0^2)$.

The scale dependence of $\kappa_{4K}$ can most easily be derived
by observing that this parameter is related to
$a_1^K$ and quark masses by the equations of motion \cite{BL04}:
\begin{equation}
\kappa_{4K} = -\frac{1}{8}\,\frac{m_s-m_q}{m_s+m_q} - \frac{9}{40}\,
a_1^K + \frac{m_s^2-m_q^2}{2m_K^2}.
\label{eq:BL04}
\end{equation}
Taking into account the known scale dependence
of $a_1^K$ and $m_{s,q}$, one obtains
\begin{equation}
\kappa_{4K}(\mu^2) = \kappa_{4K}(\mu_0^2) -
\frac{9}{40}\,\left(L^{32/(9\beta_0)} - 1\right) a_1^K(\mu_0^2) +
\left(L^{8/\beta_0} - 1\right)
\frac{[m_s^2-m_q^2](\mu_0^2)}{2m_K^2}\,.
\end{equation}

To NLO in conformal spin, the discussion becomes more involved. 
As explained in Ref.~\cite{BF90}, for massless quarks the 
corresponding contributions can be expressed in terms of matrix 
elements of the three existing G-parity-even local quark--antiquark--gluon 
operators of twist-4. These three operators are not independent, however,
but related by the QCD equations of motion. One is left with 
one new non-perturbative parameter only, call it $\omega_{4K}$,%
\footnote{In the notation of Ref.~\cite{BF90} $ \omega_{4} = (8/21)\epsilon$.}
which can be defined as     
\begin{eqnarray}
\lefteqn{\langle 0 | \bar q [iD_\mu,ig\widetilde{G}_{\nu\xi}] \gamma_\xi s -
\frac{4}{9}\, i\partial_\mu \bar q i g
\widetilde{G}_{\nu\xi}\gamma_\xi s | K(P)\rangle\
=}\hspace*{0.5cm}\nonumber\\
& = & f_K\delta_K^2 \omega_{4K} \left(P_\mu P_\nu - \frac{1}{4}\, m_K^2
g_{\mu\nu}\right) + {\mathcal O}({\rm twist\ 5}).
\end{eqnarray}
The scale dependence of $\omega_{4K}$, for massless quarks, is given by
$$
[\delta_K^2\omega_{4K}](\mu^2) =
L^{10/\beta_0}\, 
[\delta_K^2\omega_{4K}](\mu_0^2)\,.
$$
{}For massive quarks, a distinction must be made between G-parity-conserving and
G-pa\-ri\-ty-brea\-king contributions. G-parity-conserving corrections do not 
involve new operators, and the difference to the massless case is mainly due 
to corrections proportional to the meson mass. 
This case is described in detail in Refs.~\cite{BBKT,PB98}.
Here we just quote the results obtained in Ref.~\cite{PB98}: 
\begin{equation}\label{eq:T4rels}
\begin{array}[b]{@{}r@{\:=\:}l@{\qquad}r@{\:=\:}l@{}}
\phi_1^K & \ds\frac{21}{8}\,\delta_K^2\omega_{4K}
-\frac{9}{20}\, m_K^2 a_2^K , &
\wt{\phi}^K_2 & \ds \frac{21}{8}\,\delta_K^2 \omega_{4K},\\[15pt]
\psi_1^K & \ds\frac{7}{4}\,\delta_K^2 \omega_{4K} -\frac{3}{20}\,  m_K^2 a_2^K,&
\psi_2^K & \ds\frac{7}{2}\,\delta_K^2\omega_{4K} + \frac{3}{20}\,
m_K^2  a_2^K.\\[-5pt]
\end{array}
\end{equation}
The G-parity-breaking contributions, on the other hand, involve a different set 
of local operators and in particular 
$$\bar q \gamma_z\gamma_5 D_{\xi}gG^{\xi z} s 
   = -g^2 \sum_{\psi=u,d,s} (\bar q \gamma_z\gamma_5 t^a s)  (\bar \psi \gamma_z t^a \psi)$$
which determines the normalization and the leading conformal spin contribution 
to the DA $\Xi_{4;K}(\underline{\alpha})$ defined in
   Eq.~(\ref{Xi}). Hence, a complete treatment of 
G-parity-breaking corrections to twist-4 DAs requires also the inclusion of
  $\Xi_{4;K}$.

 It is beyond the scope of this paper to work out the corresponding
 relations between the matrix elements of local 
operators and expansion coefficients. For this reason, and also because QCD sum-rule
estimates of matrix elements of large mass dimension are not very reliable, we
adopt a different approach and estimate G-parity-breaking corrections
 of spin $J=4$
using the renormalon model of Ref.~\cite{BGG04}. The general idea of this technique is 
to estimate matrix elements of ``genuine'' twist-4 operators by the quadratically divergent 
contributions that appear when the matrix elements are defined using a
hard UV cut-off, see
Ref.~\cite{BGG04} for details and further references. In this way, three-particle twist-4 DAs
can be expressed in terms of the leading-twist DA $\phi_2$:
\begin{eqnarray}
\label{RMthree}
\Psi_{4;K}^{\rm ren}(\alpha_1,\alpha_2,\alpha_3)&=&
\phantom{-}\frac{\delta^2_K}{6}\left[
\frac{\phi_{2;K}(\alpha_1)}{1-\alpha_1}-\frac{\phi_{2;K}(\bar\alpha_2)}{1-\alpha_2}
\right],
\nonumber \\
\Phi_{4;K}^{\rm ren}(\alpha_1,\alpha_2,\alpha_3)&=&\phantom{-}\frac{\delta_K^2}{3}\left[
\frac{\alpha_2\phi_{2;K}(\alpha_1)}{(1-\alpha_1)^2}
-\frac{\alpha_1\phi_{2;K}(\bar\alpha_2)}{(1-\alpha_2)^2}
\right],
\nonumber\\
\widetilde\Psi_{4;K}^{\rm ren}(\alpha_1,\alpha_2,\alpha_3)&=&
\phantom{-}\frac{\delta_K^2}{6}\left[
\frac{\phi_{2;K}(\alpha_1)}{1-\alpha_1}+\frac{\phi_{2;K}(\bar\alpha_2)}{1-\alpha_2}
\right],
\nonumber \\
\widetilde\Phi_{4;K}^{\rm ren}(\alpha_1,\alpha_2,\alpha_3)&=&-\frac{\delta_K^2}{3}\left[
\frac{\alpha_2\phi_{2;K}(\alpha_1)}{(1-\alpha_1)^2}
+\frac{\alpha_1\phi_{2;K}(\bar\alpha_2)}{(1-\alpha_2)^2}
\right],
\nonumber\\
\Xi_{2;K}^{\rm ren}(\alpha_1,\alpha_2,\alpha_3)&=&
-\frac{2\delta^2_K}{3}
\left[\frac{\alpha_2\,\phi_{2;K}(\alpha_1)}{1-\alpha_1}
-\frac{\alpha_1\,\phi_{2;K}(\bar\alpha_2)}{1-\alpha_2}\right],
\end{eqnarray}
where, in difference to \cite{BGG04}, we do not assume that
$\phi_2(u)$ is symmetric under the exchange $u\leftrightarrow 1-u$.

The expressions in (\ref{RMthree}) do not rely on conformal expansion
and contain the contributions of all conformal partial waves. 
Projecting onto the contributions with the lowest spin $J=3,4$, we obtain
\begin{equation}
\renewcommand{\arraystretch}{2.2}
\begin{array}[b]{l@{\quad}l@{\quad}l}
\ds\phi_0^K = 0 & \ds\phi_1^K = \frac{7}{12}\delta^2_K & 
\ds\phi_2^K =  -\frac{7}{20}a_1^K\delta^2_K, \\
\ds\widetilde\phi_0^K = -\frac{1}{3}\delta^2_K, & 
\ds\widetilde\phi_1^K = -\frac{7}{4}a_1^K\delta^2_K, & 
\ds\widetilde\phi_2^K =  \frac{7}{12}\delta^2_K, \\
\ds\psi_0^K = -\frac{1}{3}\delta^2_K, & \ds\psi_1^K =
\frac{7}{18}\delta^2_K, & \ds\psi_2^K =  \frac{7}{9}\delta^2_K,\\
\ds\theta_0^K = 0, & \ds\theta_1^K = \frac{7}{10}a_1^K\delta^2_K,&
\ds\theta_2^K =  -\frac{7}{5}a_1^K\delta^2_K\,.
\end{array}
\label{thetas}
\renewcommand{\arraystretch}{1}
\end{equation}
It follows that in the renormalon model 
\begin{equation}
   \omega_{4K} = \omega_{4\pi} = \frac{2}{9},
\end{equation}
which is in good agreement with  direct QCD sum-rule calculations \cite{BF90}.
We also find
\begin{equation}
  \Xi^K_0 = \frac{1}{5}a_1^K \delta^2_K\,.
\end{equation}
Note that in the renormalon model $\theta_0^K = 0$. This is due to the
fact that the contribution in 
$\kappa_{4K}$ in Eq.~(\ref{eq:theta0}) is obtained as the matrix
element of the operator (\ref{eq:def33})
which vanishes by the EOM (up to a total derivative),
see Eq.~(\ref{eq:BL04}). Therefore, against appearances, this contribution has to be 
interpreted as ``kinematic'' power correction induced by the
non-vanishing $K$-meson mass rather than a ``genuine'' twist-4 effect. 

We are now  in the position to derive expressions for the
two-particle DAs of twist 4. They are defined as
\begin{eqnarray}
\langle 0 | \bar q(x)[x,-x]\gamma_\mu\gamma_5 s(-x)|K(P)\rangle
& = & i f_K P_\mu \int_0^1 du \, e^{i\xi Px} \left(\phi_{2;K}(u) +
\frac{1}{4}\,x^2 \phi_{4;K}(u)\right)\nonumber\\
&&{}+\frac{i}{2}\, f_K\, \frac{1}{Px}\, x_\mu  \int_0^1 du \, e^{i\xi
  Px} \psi_{4;K}(u),
\end{eqnarray}
which is the extension of Eq.~(\ref{eq:T2}) to twist-4
accuracy.\footnote{$\psi_{4;K}$ and $\phi_{4;K}$ are related to the
  DAs defined in Ref.~\cite{PB98} by $\phi_{4;K} = m_K^2 g_K$ and
  $\psi_{4;K} = m_K^2 {\mathbb A}_K$.}  From the
operator relations (\ref{eq:oprel1}) and (\ref{eq:oprel2}), we
obtain
\begin{eqnarray}
\psi_{K;4}(u) & = & m_K^2\{2\phi^p_{3;K}(u) - \phi_{2;K}(u)\} + \frac{d}{du}
\int_0^u d\alpha_1 \int_0^{\bar u} d\alpha_2 \,\frac{2\left(
  \Phi_{4;K}(\underline{\alpha}) - 
2 \Psi_{4;K}(\underline{\alpha})\right)}{1-\alpha_1-\alpha_2}\,,
\hspace*{0.8cm}\label{eq:gpi}\\
\frac{d^2\phi_{4;K}(u)}{du^2} & = & 12\psi_{4;K}(u)-12 m_K^2 \phi_{2;K}(u) - 
2 \frac{d}{du}\left[(2u-1) (m_K^2\phi_{2;K}(u) + \psi_{4;K}(u))\right]
\nonumber\\
& & + \frac{d^2}{du^2}\,
\int_0^u d\alpha_1 \int_0^{\bar u} d\alpha_2 \,\frac{4( 2 \Psi_{4;K}(\underline{\alpha}) -
\Phi_{4;K}(\underline{\alpha}) )}{(1-\alpha_1-\alpha_2)^2}
\, (\alpha_1-\alpha_2 - (2u-1)) \nonumber\\
&& {}+4\,\frac{m_s-m_q}{m_s+m_q}\,m_K^2\,\frac{d \phi_{3;K}^p(u)}{du}\label{A}
\end{eqnarray}
with the boundary condition $\phi_{4;K}(0) = \phi_{4;K}(1) = 0.$

We solve this relations splitting the result in ``genuine'' twist-4
contributions $\psi_{4;K}^{T4}$ and Wandzura-Wilczek-type
mass corrections $\psi_{4;K}^{WW}$ as
\begin{equation}
\psi_{4;K}(u) = \psi_{4;K}^{T4}(u) + \psi_{4;K}^{WW}(u)
\label{eq:T4psi}
\end{equation}
with
\begin{eqnarray}
\psi_{4;K}^{T4}(u) & = &  \frac{20}{3}\,\delta_K^2 C_2^{1/2}(2u-1) + 5
\left\{ 5 \theta_1^K - \theta_2^K\right\} C_3^{1/2}(2u-1)\,,\label{eq:T4psiCE}\\
\psi_{4;K}^{WW}(u) &=& m_K^2\left\{1+6 \rho^K_+ (1+6 a_2^K) - 18 \rho^K_-
a_1^K\right\} C_0^{1/2}(2u-1)
\nonumber\\
&&{} + m_K^2 \left\{ -12 \kappa_{4K} -
\frac{9}{5}\,a_1^K + 27 \rho^K_+ a_1^K - 3 \rho^K_-(1+18 a_2^K)\right\}C_1^{1/2}(2u-1)
\nonumber\\
&& {}+ \left\{m_K^2\left(1 + \frac{18}{7}\,a_2^K + 30\rho^K_+ a_2^K - 6
\rho^K_- a_1^K\right) + 60\,\frac{f_{3K}}{f_K}\,(m_s+m_q)\right\}C_2^{1/2}(2u-1)
\nonumber\\
&&{}+\left\{m_K^2\left(\frac{9}{5}\,a_1^K +
\frac{16}{3}\,\kappa_{4K}-9 \rho^K_- a_2^K\right) + 20\,
\frac{f_{3K}}{f_K}\,(m_s+m_q) \lambda_{3K}\right\}
C_3^{1/2}(2u-1)
\nonumber\\
&&{}+\left\{-\frac{9}{28}\, m_K^2 a_2^K - 6\,
\frac{f_{3K}}{f_K}\,(m_s+m_q)\omega_{3K}\right\}
C_4^{1/2}(2u-1)
\nonumber\\
&&{}+ 6 m_q (m_s+m_q) (1+3a_1^K+6 a_2^K) \ln \bar u + 6 m_s (m_s+m_q)
(1-3a_1^K+6 a_2^K) \ln u\,,\nonumber\\[-12pt]
\label{eq:phi4}
\end{eqnarray} 
where $\xi = 2u-1$, see Eq.~(\ref{eq:xi}). $\psi_{4;K}^{WW}$ vanishes
for $m_K\to 0$ and $m_{s,q}\to 0$.

The complete expression for 
\begin{equation}
\phi_{4;K}(u) = \phi_{4;K}^{T4}(u) + \phi_{4;K}^{WW}(u)
\label{eq:T4phi}
\end{equation}
is rather lengthy. We find for the ``genuine'' twist-4 part: 
\begin{eqnarray}
\phi_{4;K}^{T4}(u) & = & \frac{200}{3}\,\delta_K^2 u^2\bar u^2 + 20 u^2\bar
u^2 \xi \left\{ 4 \theta_1^K - 5 \theta_2^K\right\}
\nonumber\\
&&{} + 21
\delta_K^2 \omega_{4K} \Big\{ u\bar u (2+13 u\bar u) + \left[ 2 u^3(6
  u^2-15u+10)\ln u\right] + [u\leftrightarrow\bar
  u]\Big\}
\nonumber\\
&&{}+40 \phi_2^K \left\{ u\bar u \xi (2-3 u\bar u) - \left[ 2
  u^3(u-2)\ln u\right] + \left[u\leftrightarrow \bar u\right]\right\},
\label{eq:T4phiCE}
\end{eqnarray}
and for the mass-corrections, neglecting numerically small terms of order $m_s^2$: 
\begin{eqnarray}  
\phi_{4;K}^{WW}(u) & = &\frac{16}{3}\,m_K^2\kappa_{4K}\Big\{ u\bar u\xi (1-2u\bar u) +
\left[5(u-2)u^3 \ln u\right] - [u\leftrightarrow\bar u
]\Big\}
\nonumber\\
&&{}+4\,\frac{f_{3K}}{f_K}\,(m_s+m_q) u\bar u \left\{30\left( 1-
\xi\,\frac{m_s-m_q}{m_s+m_q}\right)\right.
\nonumber\\
&&{} \hspace*{1cm}+ 10\lambda_{3K} \left(\xi
\left[1-u\bar u\right]-\frac{m_s-m_q}{m_s+m_q}\,[1-5u\bar u]\right)
\nonumber\\
&&{}\hspace*{1cm}\left. - \omega_{3K} \left( 3-21 u\bar u +28 u^2\bar u^2 +
3 \xi \frac{m_s-m_q}{m_s+m_q}[1-7u\bar u]\right)\right\}
\nonumber\\
&&{}-\frac{36}{5}m_K^2 a_2^K \Big\{
\frac{1}{4} u\bar u (4-9u\bar u +110 u^2\bar u^2) +
 [u^3(10-15u+6u^2)\ln u] +
[u\leftrightarrow\bar u]\Big\}
\nonumber\\
&&{}+4 m_K^2 \, u \bar u\, (1+3u\bar u)\left(1+\frac{9}{5}a_1^K \,\xi\right)\,.
\end{eqnarray}
The DAs for  $\bar K$
mesons are obtained by replacing $u$ by $1-u$. Note that $\psi_{4;K}$
has logarithmic end-point singularities for finite quark mass, whereas
$\phi_{4;K}$ has no such singularities, so that one can safely neglect
the ${\mathcal O}(m_s^2)$ terms.

The expressions given above provide a self-consistent model of the twist-4 DAs 
which includes the first three terms of the conformal expansion.%
\footnote{One shortcoming of the model is 
that G-parity-breaking meson mass corrections of spin $J=4$ are missing and 
we only include the ``genuine'' G-parity-breaking twist-4 corrections estimated 
in the renormalon model. Numerically, both effects may be of the same order.} 
An estimate of the 
contribution of higher orders can be obtained using the renormalon model.
In this case, the ``genuine'' twist-4 contributions to the 
two-particle DAs given in Eqs.~(\ref{eq:T4psi}) and (\ref{eq:T4phi}) have to 
be replaced by     
\begin{eqnarray}
\label{RMtwo}
 \phi_{4;K}^{T4,{\rm ren}}(u)
 &=&\frac{8}{3}\,\delta_K^2\int_0^1\!\!dv\,\phi_{2;K}(v)\bigg\{
 \frac{1}{v^2}\,\bigg[u^2+u+(v-u)\ln\left(1-\frac{u}{v}\right)\bigg] \theta(v-u)\nonumber
  \\ &&\hspace*{90pt} +\,
 \frac{1}{\bar v^2}\,\bigg[\bar 
u^2 + \bar u+(u-v)\ln\left(1-\frac{\bar u}{\bar v}\right)\bigg] \theta(u-v)\bigg\},
 \nonumber \\ 
 \psi_{4;K}^{T4,{\rm ren}}(u)
&=&\frac{\delta_K^2}{3}\,\frac{d^2}{du^2}\int_0^1dv\,\phi_{2;K}( v)\left\{
\left(\frac{u}{v}\right)^2 \theta(v-u)+
\left(\frac{\bar u}{\bar v}\right)^2 \theta(u-v)\right\}
\end{eqnarray}
and used in combination with the complete renormalon-model expressions for the three-particle
DAs given in Eq.~(\ref{RMthree}). As explained in Ref.~\cite{BGG04}, the renormalon model 
does not take into account the damping of higher conformal-spin contributions by the increasing 
anomalous dimensions and, therefore, provides an upper bound for their contribution.
The effect of these corrections is, most importantly, to significantly enhance 
the end-point behaviour of higher-twist DAs in some cases, which can
be important in phenomenological applications.

\section{Models for Distribution Amplitudes}\label{sec:5}
\setcounter{equation}{0}

In this section we compile the numerical estimates of all necessary parameters 
and present explicit models of
the twist-3 and -4 two-particle
distribution amplitudes that we introduced in Sections~\ref{sec:3} and \ref{sec:4}.
The important point is that these DAs are related to three-particle ones by exact
QCD equations of motion and have to be used together; this
guarantees the consistency of the approximation.
Our approximation thus introduces a minimum number of non-perturbative 
parameters, which are defined as  matrix elements of certain local operators 
between the vacuum and the meson state, and which we estimate using QCD sum rules. More
sophisticated models can be constructed in a systematic way by adding 
contributions of higher conformal partial waves when estimates of the relevant 
non-perturbative matrix elements will become available. 

Our  approach involves the implicit assumption that the conformal partial 
wave expansion is well convergent. This can be justified rigorously 
at large scales, since the anomalous dimensions of all involved 
operators increase logarithmically with the conformal spin $J$, but 
is non-trivial at relatively low scales of order $\mu \sim (1$--$2)\,$GeV
which we choose as reference scale. An upper bound for the contribution
of higher partial waves can be obtained from the renormalon model.
%

Since orthogonal polynomials of high orders are rapidly oscillating 
functions, a truncated expansion in conformal partial waves is, almost necessarily,
oscillatory as well. Such a behaviour is clearly unphysical,
but this does not constitute a real problem since  physical observables 
are given by convolution integrals of distribution amplitudes with
smooth coefficient functions. 
A classical example for this
feature is the $\gamma\gamma^*$-meson form factor, which is governed
by the quantity
$$
\int du\,\frac{1}{u}\, \phi(u) \sim \sum a_i,
$$
where the coefficients $a_i$ are exactly the ``reduced matrix elements''
in the conformal expansion.
The oscillating terms are averaged over and strongly suppressed. 
Stated otherwise: 
models of distribution amplitudes should generally be understood as
distributions (in the mathematical sense).    

\begin{table}
\renewcommand{\arraystretch}{1.2}\addtolength{\arraycolsep}{1pt}
$$
\begin{array}{l|l|l||l|l|l|l}
\hline
K & \mu = 1\,{\rm GeV} &  \mu = 2\,{\rm GeV} & \pi & \mu = 1\,{\rm GeV} &
\mu = 2\,{\rm GeV} & {\rm Remarks}\\\hline
\overline{m}_s & 137\pm 27 & 100\pm 20 &  
\overline{m}_q & 5.6\pm1.6 & 4.1\pm1.1 & \mbox{in units of MeV; see\ Sec.~\ref{sec:2}}\\ 
a_1^K & 0.06\pm 0.03 & 0.05\pm 0.02 & a_1^\pi & 0 & 0 & \mbox{taken from
  Ref.~\cite{BZ05}; G-odd}\\
a_2^K & 0.25\pm 0.15 & 0.17\pm 0.10 & a_2^\pi & 0.25\pm 0.15 & 0.17\pm
0.10 & \mbox{$SU(3)$ breaking small;}\\[-3pt]
& & & & & & \mbox{average over various Refs.}\\
f_{3K} & 0.45\pm 0.15 & 0.33\pm0.11 & f_{3\pi} & 0.45\pm 0.15 &
0.31\pm 0.10 & \mbox{in units of $10^{-2}\,$GeV$^2$}\\
\omega_{3K} & -1.2\pm 0.7& -0.9\pm 0.5 & \omega_{3\pi} & -1.5\pm0.7 &
-1.1\pm 0.5 \\
\lambda_{3K} & 1.6\pm 0.4 & 1.45\pm 0.35 & \lambda_{3\pi} & 0 & 0 & \mbox{G-odd}\\
\delta_K^2 & 0.20\pm0.06 & 0.17\pm0.05 & \delta_{\pi}^2 & 0.18\pm0.06
& 0.14\pm0.05 & \mbox{in units of GeV$^2$}\\
\kappa_{4K} & -0.09\pm 0.02 & -0.10 \pm 0.02 & \kappa_{4\pi} & 0 & 0 & \mbox{taken from
  Ref.~\cite{BZ06}; G-odd}\\
\omega_{4K} & 0.2\pm 0.1 & 0.13\pm 0.07 & \omega_{4\pi} & 0.2\pm 0.1 &
0.13\pm 0.07 & \mbox{taken from
  Ref.~\cite{BF90};}\\[-3pt]
&&&&&& \mbox{$SU(3)$ breaking not incl.} \\\hline
\end{array}
$$
\renewcommand{\arraystretch}{1}\addtolength{\arraycolsep}{-1pt}
\caption[]{\sf Hadronic parameters for the $K$ DAs. We also give the
  corresponding parameters for the $\pi$, which are a by-product of our
  calculations. All parameters have been calculated in this paper
  at $\mu=1\,$GeV, unless stated otherwise. The evolution between 1 and
  2$\,$GeV is done at NLO accuracy for $\overline{m}_{q,s}$ and
  $a_{1,2}^{\pi,K}$, and at LO accuracy for the other parameters. The
  twist-4 parameters $\theta_i^K$, $\phi^K_i$ etc.\ are given by
  Eq.~(\ref{thetas}), based on the renormalon model.}\label{tab:num}
\end{table}
\begin{figure}[p]
$$\epsfxsize=0.47\textwidth\epsffile{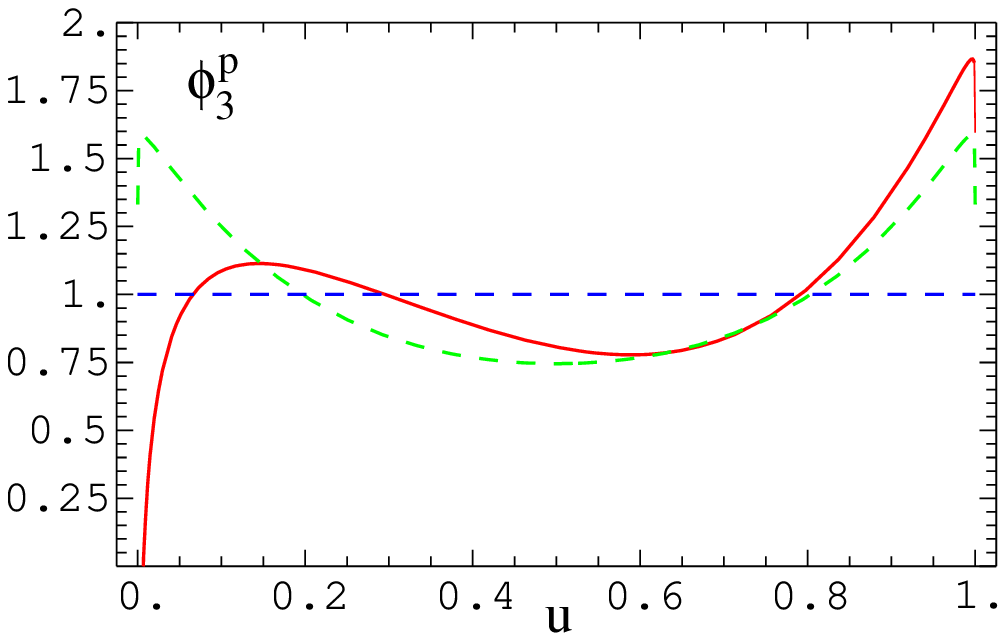}\quad
\epsfxsize=0.47\textwidth\epsffile{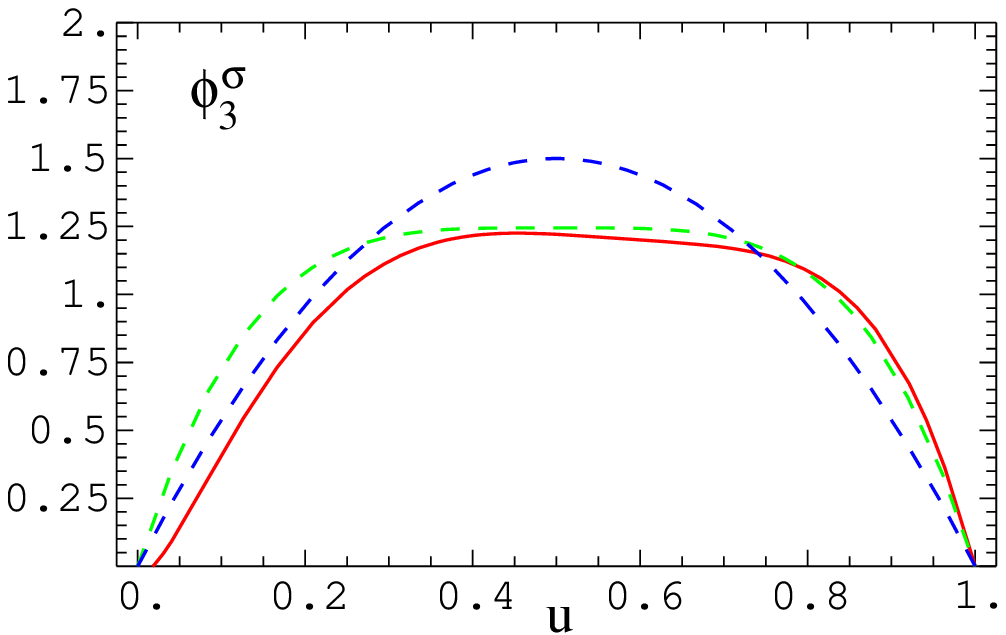}$$
\vspace*{-15pt}
\caption{\sf Left panel: $\phi_{3}^p$ as a function of $u$ for the central
  value of the hadronic parameters, for $\mu=1\,$GeV. Red (solid) line: $\phi_{3;K}^p$,
  green (long dashed): $\phi_{3;\pi}^\sigma$, blue (short dashed):
  asymptotic DA. Right panel: same for $\phi_{3}^\sigma$.}\label{fig:1}
$$\epsfxsize=0.47\textwidth\epsffile{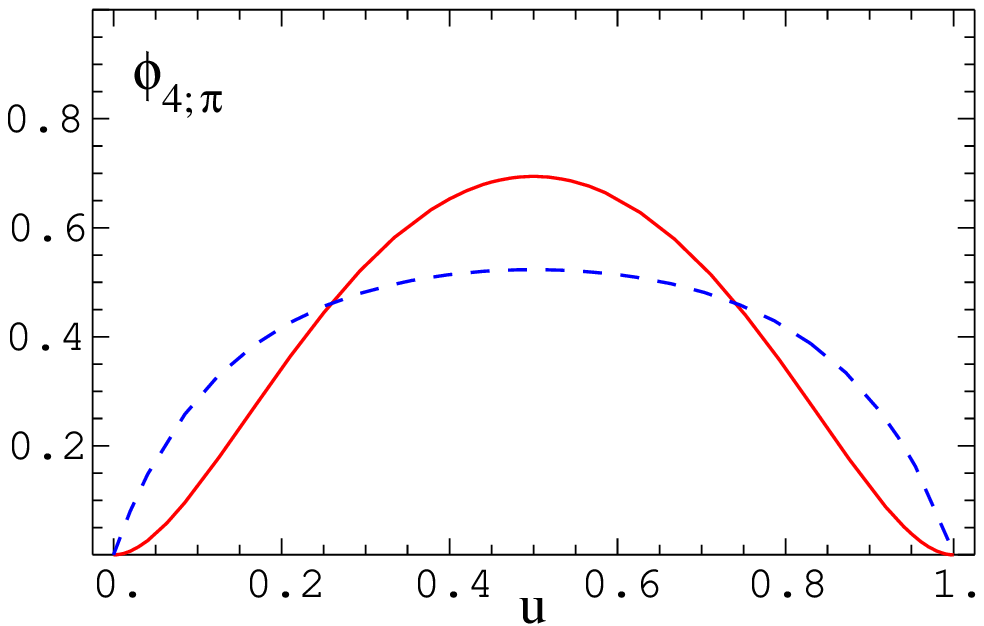}\quad
\epsfxsize=0.47\textwidth\epsffile{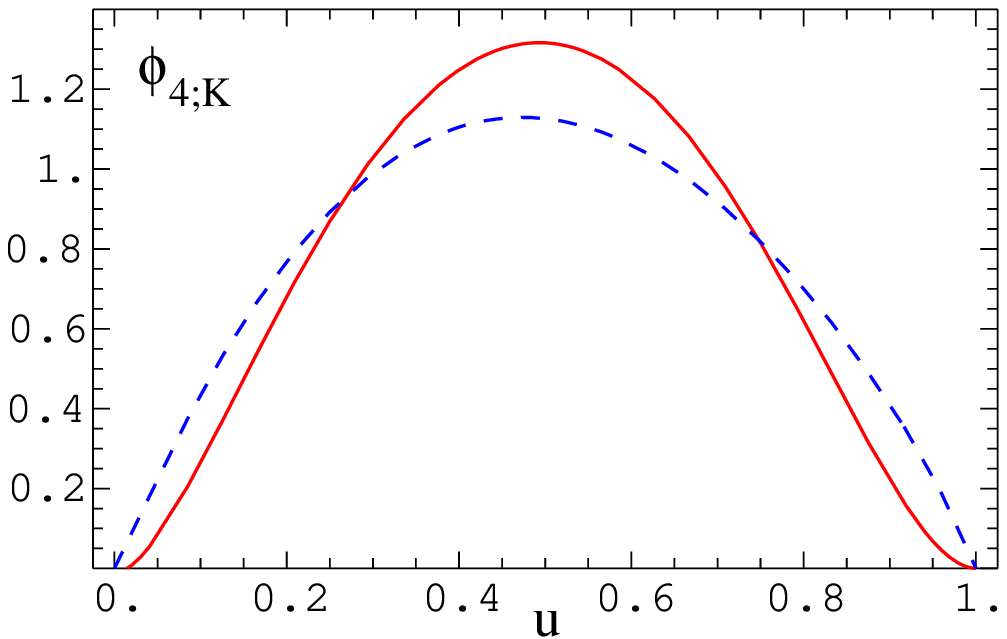}$$
\vspace*{-15pt}
\caption{\sf Left panel: $\phi_{4;\pi}$ as a function of $u$ for the central
  value of the hadronic parameters, for $\mu=1\,$GeV. Red (solid)
  line: $\phi_{4;\pi}$ in conformal expansion, blue (dashed):
  $\phi_{4;\pi}$ using the renormalon model
  $\phi_{4;\pi}^{T4,\mbox{\rm\scriptsize ren}}$ for the genuine twist-4 corrections. 
Right panel: same for $\phi_{4;K}$.}\label{fig:2}
$$\epsfxsize=0.47\textwidth\epsffile{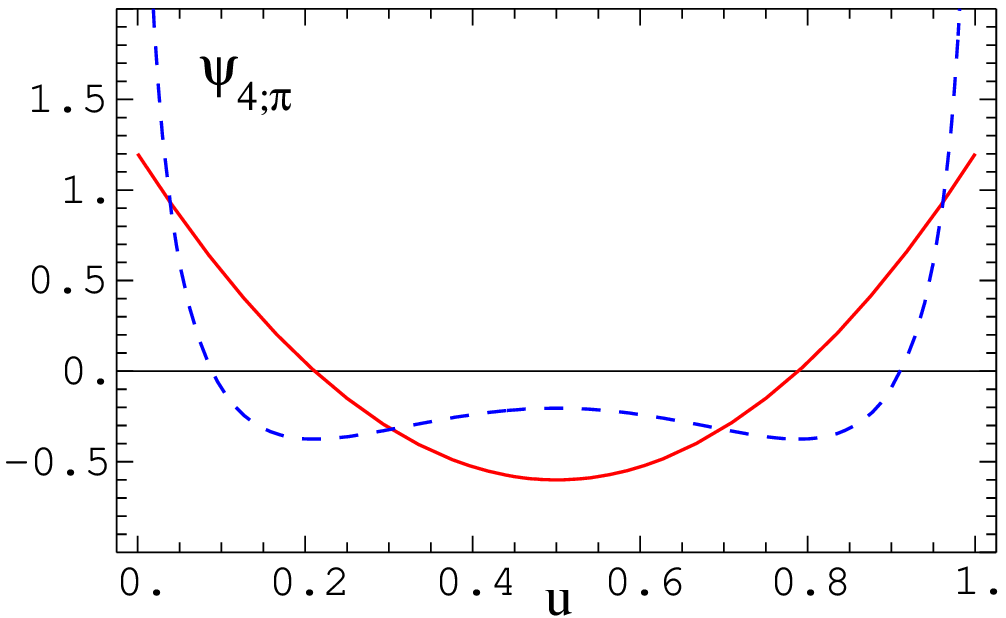}\quad
\epsfxsize=0.47\textwidth\epsffile{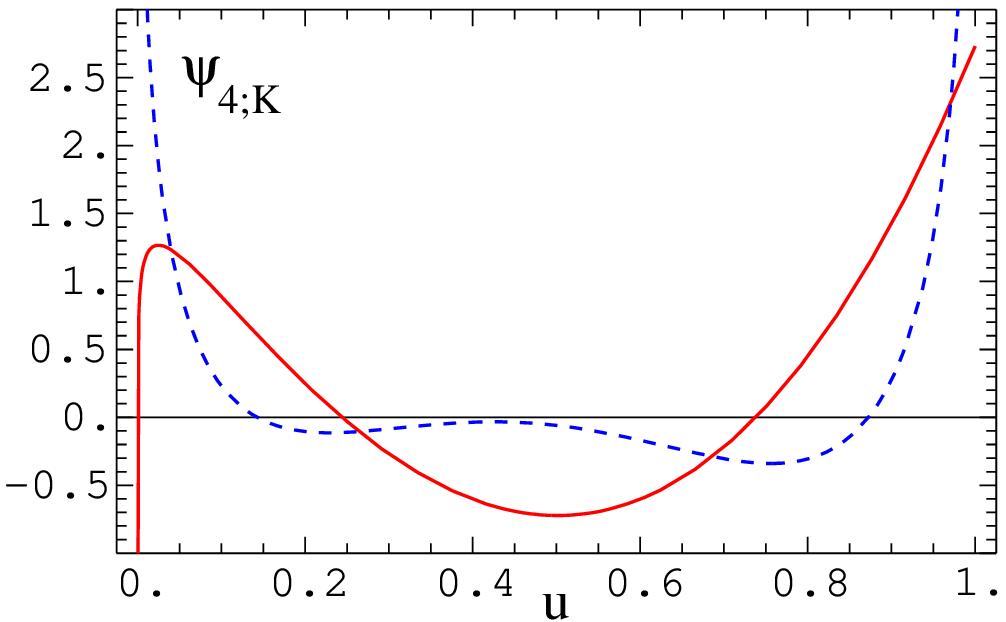}$$
\vspace*{-15pt}
\caption{\sf Same as Figure~\ref{fig:2} for $\psi_4$.}\label{fig:2a}
\end{figure}

We give all relevant numerical
input parameters for our model DAs in
Table~\ref{tab:num}, at the scale $\mu=1\,$GeV, which is appropriate for
QCD sum-rule results, and, using the LO and NLO scaling relations
given in Section~\ref{sec:3} and \ref{sec:4}, at the scale $\mu=2\,$GeV,
in order to facilitate the comparison with future lattice
determinations of these quantities. The mixing of $K$-meson parameters
with operators of lower twist depending on $m_s$ is numerically small. 

The parameters related to twist-2 matrix
elements have been determined using various methods; see the
discussion in Section~\ref{sec:2}. 
Matrix elements of twist-3 and 4
operators for the $\pi$ meson were calculated
a long time ago from QCD sum rules \cite{misuse,CZ,BF90}. 
In this paper, we perform a complete reanalysis of these sum rules 
and also include  $SU(3)$-breaking effects relevant for the $K$ meson.
The corresponding sum rules and plots  are given in the
appendices. One important result is that we cannot confirm the sum rule 
for $f_{3\pi}$ derived in Ref.~\cite{CZ} and that our numerical value is considerably larger
than that found in this paper. On the other hand, our central value for $\delta^2_\pi$ 
is similar to the one obtained in Ref.~\cite{misuse}, see also  Ref.~\cite{Bakulev:2002uc}.

Finally, in Figure~\ref{fig:1} we plot the
twist-3 and -4 two-particle DAs for the $\pi$ meson, assuming massless quarks,
and for the $K$ meson, together with the corresponding asymptotic DAs.
The figures show that quark-mass corrections significantly modify
 the end-point behaviour of $\phi_3^p$, where they induce
a logarithmic end-point divergency, even if the contributions of gluonic operators are
 neglected. This is not a problem because, as mentioned above, the DAs
 themselves need not be finite, it is only their convolution with
 perturbative scattering amplitudes that is meaningful. In
 Figures~\ref{fig:2} and \ref{fig:2a} we show the twist-4 two-particle
 DAs $\phi_4$ and $\psi_4$, also for the $\pi$ (left panels) and the
 $K$ (right panels). The solid (red) curve in Figure~\ref{fig:2} is obtained from
 Eq.~(\ref{eq:T4phi}) using the conformal expansion (\ref{eq:T4phiCE})
 to NLO in the conformal spin, whereas the dashed (blue)
 curve includes the higher-spin contributions to the genuine twist-4
 corrections as given by the renormalon model (\ref{RMtwo}). The
 mass corrections $\phi_{4;\pi}^{WW}$ vanish for the pion. It is
 clear that the higher-order contributions induced by (\ref{RMtwo})
 modify both the end-point behaviour of $\phi_{4;\pi}$ and the size of
 the DA away from the end-points. For the $K$, the absolute difference between
 both curves at, say, $u=\frac{1}{2}$, is very nearly the same as for
 $\pi$, but the relative difference is much reduced because of large
 $SU(3)$-breaking effects induced by the mass-dependent contribution
 $\phi_{4;K}^{WW}$. Also note that the asymmetry of the curves induced
 by the non-vanishing value of $a_1^K$ is not very pronounced, which
 is due to the smallness of that parameter as compared to $a_0^K=1$
 and $a_2^K$. In Figure~\ref{fig:2a} we plot $\psi_4$, with the same
 meaning of the curves as in Figure~\ref{fig:2}. Also here it is
 obvious that the renormalon model modifies the end-point behaviour of
 the DA, in particular for $\psi_{4;K}$, where it changes the sign
 of the logarithmic divergence at $u=0$. 

\section{Summary and Conclusions}\label{sec:6}
\setcounter{equation}{0}

In this paper we have studied the twist-3 and -4 two- and three-particle
distribution amplitudes of $K$-mesons in QCD and expressed them in a
model-independent way by a minimal number of non-perturbative
parameters. The work presented here is an extension of
Refs.~\cite{BF90,PB98,BGG04} 
and completes the analysis of  $SU(3)$-breaking
corrections by also including G-parity-breaking corrections in
$m_s-m_q$. Our approach consists of two components.
One is the use of the QCD
equations of motion, which allow dynamically dependent
DAs to be expressed in terms of independent ones. The other ingredient is conformal
expansion, which makes it possible to separate transverse and
longitudinal variables in the wave functions, the former ones being
governed by renormalization-group equations, the latter ones being
described in terms of irreducible representations of the corresponding
symmetry group.
We have derived expressions for all twist-3 and -4 two- and
three-particle distribution amplitudes to next-to-leading order in the
conformal expansion, including both chiral corrections ${\mathcal
  O}(m_s+m_q)$ and G-parity-breaking corrections ${\mathcal
  O}(m_s-m_q)$; the corresponding formulas are given in
Secs.~\ref{sec:3} and \ref{sec:4}.
We have also generalized the renormalon model of Ref.~\cite{BGG04} to describe
$SU(3)$-breaking contributions to high-order conformal partial waves.

We have done a complete reanalysis of the numerical values of the
relevant higher-twist hadronic parameters from QCD sum rules. 
Our sum rules can be compared, in the chiral limit, with
existing calculations for the $\pi$ \cite{misuse,CZ}. We confirm the
sum rule for the twist-4 matrix element $\delta^2_\pi$ quoted in
Ref.~\cite{misuse}, but obtain different results for the twist-3 matrix
elements given in Ref.~\cite{CZ}, which lead to a 50\% increase in the
numerical value of the coupling $f_{3\pi}$. Whenever possible, we have aimed at
determining these matrix elements from more than one sum rule; we
find mutually consistent results, which provides a consistency check
of the approach. We have also studied the scale-dependence of all
parameters to leading-logarithmic, or, if possible,
next-to-leading-logarithmic accuracy, taking into account the mixing with operators
depending on the strange-quark mass $m_s$. 
Our final numerical results, at the scales 1 and 2~GeV, are collected
in Table~\ref{tab:num}. 


We hope that our results will contribute to a better understanding of
$SU(3)$-breaking effects in hard exclusive processes and in particular
in the decays of $B$ and $B_s$ mesons into final states containing $K$ mesons.

\section*{Acknowledgements}

P.B.\ is grateful to the University of Regensburg for hospitality.

\appendix

\section*{Appendices}
\renewcommand{\theequation}{\Alph{section}.\arabic{equation}}
\renewcommand{\thetable}{\Alph{table}}
\setcounter{section}{0}
\setcounter{table}{0}

\section{Non-Local Operator Identities}\label{app:B}
\setcounter{equation}{0}

For completeness, we quote the following non-local operator identities
from Ref.~\cite{BBKT}:
\begin{eqnarray}
\frac{\partial}{\partial x_\mu}\, \bar q(x)\gamma_\mu\gamma_5 s(-x)
& = &{} - i \int_{-1}^1 dv\, v \bar q (x) x_\alpha
gG_{\alpha\mu}(vx) \gamma_\mu\gamma_5 s(-x)\nonumber\\
& &  + (m_q-m_s) \bar q(x) i\gamma_5 s(-x),\label{eq:oprel1}\\
\partial_\mu \{\bar q(x)\gamma_\mu\gamma_5 s(-x)\}
& = & {} - i\int_{-1}^1 dv\, \bar q(x) x_\alpha
gG_{\alpha\mu}(vx) \gamma_\mu\gamma_5 s(-x)\nonumber\\
& & {} + (m_s+m_q)\bar q(x)
i\gamma_5 s(-x),\label{eq:oprel2}\\
\partial_\mu \bar q(x) \sigma_{\mu\nu}\gamma_5 s(-x) & = &
-i\,\frac{\partial\phantom{x_\nu}}{\partial x_\nu} \,\bar q(x)\gamma_5
s(-x) + \int_{-1}^1 dv\, v \bar q(x)
x_\rho gG_{\rho\nu}(vx)\gamma_5s(-x)\nonumber\\
& & {} -i\int_{-1}^1 dv\, \bar q(x) x_\rho gG_{\rho\mu}(vx)
\sigma_{\mu\nu} \gamma_5s(-x)\nonumber\\
& &  + (m_s-m_q) \bar q(x) \gamma_\nu\gamma_5 s(-x),\label{eq:oprel3}\\
\frac{\partial\phantom{x_\nu}}{\partial x_\mu} \,\bar q(x)
\sigma_{\mu\nu}\gamma_5 s(-x) & = & -i\partial_\nu \bar q(x)
\gamma_5s(-x) +
\int_{-1}^1 dv\, \bar q(x) x_\rho gG_{\rho\nu}(vx)\gamma_5s(-x)\nonumber\\
& & {} -
i\int_{-1}^1 dv\, v \bar q(x) x_\rho gG_{\rho\mu}(vx)
\sigma_{\mu\nu}\gamma_5 s(-x)\nonumber\\
& & {}- (m_s+m_q) \bar q(x) \gamma_\nu \gamma_5 s(-x).
\label{eq:oprel4}
\end{eqnarray}
Here $\partial_\mu$ is the total derivative defined as
$$
\partial_\mu \left\{ \bar q(x)\Gamma s(-x)\right\} \equiv
\left.\frac{\partial}{\partial y_\mu}\,\left\{ \bar q(x+y) [x+y,-x+y]
    \Gamma s(-x+y)\right\}\right|_{y\to 0}.
$$
By taking matrix elements of the above relations between the vacuum
and the meson state, one obtains exact integral
representations for those DAs that are not dynamically independent.

\section{Sum Rules for Twist-2 Matrix Elements}\label{app:C}
\setcounter{equation}{0}

\begin{figure}
$$\epsfxsize=0.47\textwidth\epsffile{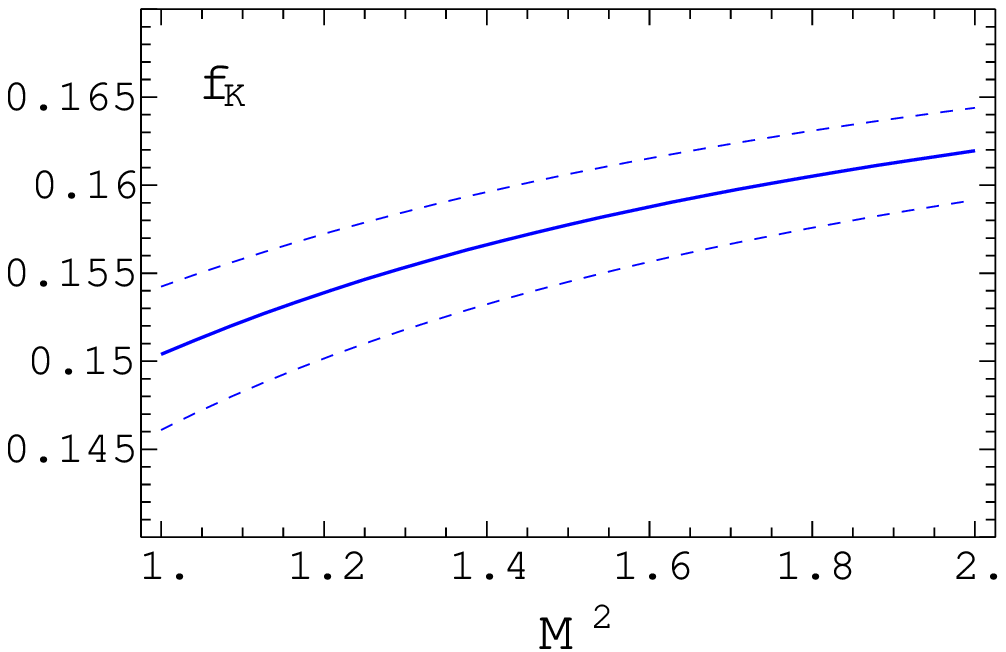}\quad
\epsfxsize=0.45\textwidth\epsffile{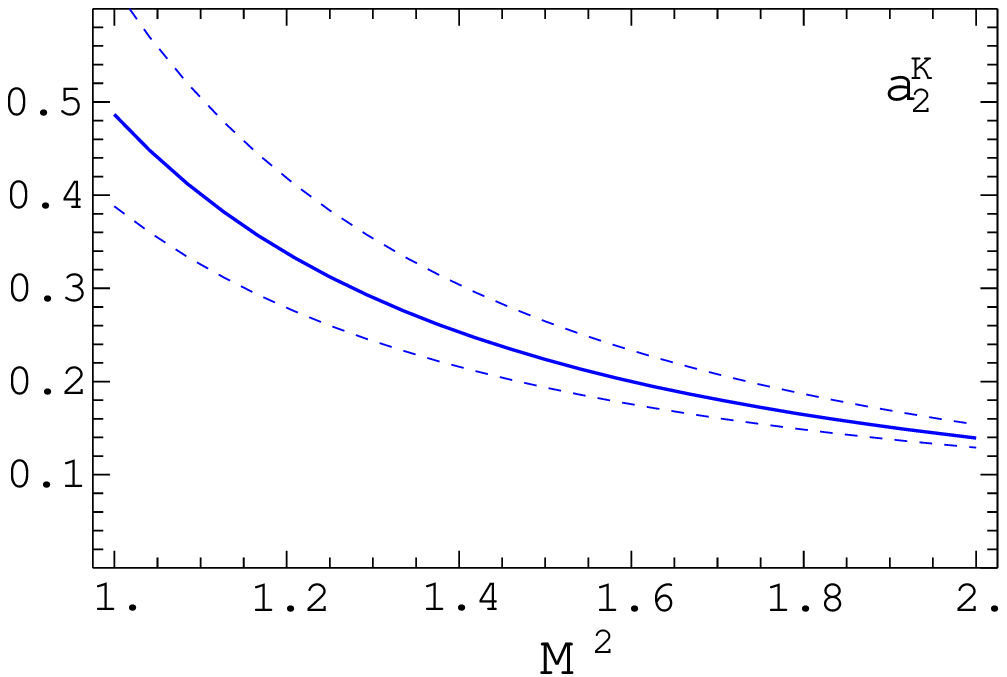}$$
\vskip-10pt
\caption[]{\sf Left panel: $f_K$ as function of the Borel parameter
  $M^2$ for $s_0=1.1\,{\rm GeV}^2$. Solid line: central values of
  input parameters, dashed lines: variation of $f_K$ within the
  allowed range of input parameters. Figure taken from
  Ref.~\cite{BZ05}. Right panel: same for $a_2^K$ at the scale $\mu=1\,$GeV. 
The results for $a_2^K$ are new.}\label{fig:app1}
\end{figure}

In this appendix we list and evaluate the QCD sum rules for twist-2
matrix elements of the $K$. 
The sum rule for $f_K$, including $SU(3)$-breaking corrections, was
calculated in Refs.~\cite{govaerts,BZ05}, that for $a_1^K$ in
Refs.~\cite{KMM04,BZ05}, and that for $a_2^K$ in Ref.~\cite{BB03},
apart from the perturbative terms in $m_s^2$ and the 
radiative corrections to the quark condensate, which are new. The sum
rules read:
\begin{eqnarray}
f_K^2 e^{-m_K^2/M^2} 
& = & \frac{1}{4\pi^2}\int\limits_{m_s^2}^{s_0}
ds\,e^{-s/M^2} \,\frac{(s-m_s^2)^2 (s+2m_s^2)}{s^3} +
\frac{\alpha_s}{\pi}\, \frac{M^2}{4\pi^2}\left( 1 -
e^{-s_0/M^2}\right)\nonumber\\
&&{} +\frac{m_s\langle \bar s s\rangle}{M^2}\left(1+\frac{m_s^2}{3M^2} + 
\frac{13}{9}\,\frac{\alpha_s}{\pi}\right)+\frac{1}{12M^2}\,
\left\langle\frac{\alpha_s}{\pi}\,G^2\right\rangle 
\left( 1+ \frac{1}{3}\,\frac{m_s^2}{M^2}\right) \nonumber\\
&&{}  +\frac{4}{3}\,\frac{\alpha_s}{\pi} \, \frac{m_s\langle \bar q
  q\rangle}{M^2}+\frac{16\pi\alpha_s}{9M^4}\,
\langle \bar q q\rangle\langle \bar s s\rangle +
\frac{16\pi\alpha_s}{81M^4}\,\left( \langle \bar q q\rangle^2 +
\langle \bar s s\rangle^2 \right),
\end{eqnarray}
\begin{eqnarray}
\lefteqn{a_1^K f_K^2 e^{-m_K^2/M^2} = 
\frac{5}{4\pi^2}\,m_s^4\int\limits_{m_s^2}^{s_0}
ds\,e^{-s/M^2} \,\frac{(s-m_s^2)^2}{s^4}}\nonumber\\
\hskip-10pt&&
{}+\frac{5m_s^2}{18M^4}\,\left\langle\frac{\alpha_s}{\pi}\,G^2\right\rangle 
\left( -\frac{1}{2}+ \gamma_E - {\rm
  Ei}\left(-\frac{s_0}{M^2}\right)+ \ln\,\frac{m_s^2}{M^2} + 
\frac{M^2}{s_0}\left( \frac{M^2}{s_0} -1\right) e^{-s_0/M^2}\right)\nonumber\\
&&{}
-\frac{5}{3}\,\frac{m_s\langle \bar s
  s\rangle}{M^2}\left\{1 + 
\frac{\alpha_s}{\pi}\left[ -\frac{124}{27} +
\frac{8}{9} \left(1-\gamma_E + \ln\,\frac{M^2}{\mu^2} +
\frac{M^2}{s_0}\,e^{-s_0/M^2} + {\rm Ei}\left(-\frac{s_0}{M^2}\right)
\right)\right]\right\}  \nonumber\\
&&{}-\frac{5}{3}\,\frac{m_s^3\squark}{M^4}-\frac{20}{27}\,\frac{\alpha_s}{\pi} \, \frac{m_s\langle \bar q
 q\rangle}{M^2} + \frac{5}{9}\, \frac{m_s\langle\bar s \sigma g Gs\rangle}{M^4}
+\frac{80\pi\alpha_s}{81M^4}\,\left( \langle \bar q q\rangle^2 -
\langle \bar s s\rangle^2 \right),\label{eq:SRa1}
\end{eqnarray}
\begin{eqnarray}
\lefteqn{a_2^K f_K^2 e^{-m_K^2/M^2} = }\nonumber\\
&&{}\frac{7}{4\pi^2}\,m_s^4\int\limits_{m_s^2}^{s_0}
ds\,e^{-s/M^2} \,\frac{(s-m_s^2)^2(2m_s^2-s)}{s^5} +
\frac{7}{72\pi^2}\,\frac{\alpha_s}{\pi} \,M^2 (1-e^{-s_0/M^2})
+\frac{7}{36M^2}\,\left\langle\frac{\alpha_s}{\pi}\,G^2\right\rangle \nonumber\\
&&{}+\frac{7}{3}\,\frac{m_s\langle \bar s
  s\rangle}{M^2}\left\{1+ 
\frac{\alpha_s}{\pi}\left[ -\frac{184}{27} +
\frac{25}{18} \left(1-\gamma_E + \ln\,\frac{M^2}{\mu^2} +
\frac{M^2}{s_0}\,e^{-s_0/M^2} + {\rm Ei}\left(-\frac{s_0}{M^2}\right)
\right)\right]\right\}  \nonumber\\
&&{}-\frac{49}{27}\,\frac{\alpha_s}{\pi} \, \frac{m_s\langle \bar q
 q\rangle}{M^2} - \frac{35}{18}\, 
\frac{m_s\langle\bar s \sigma g Gs\rangle}{M^4}
+\frac{224\pi\alpha_s}{81M^4}\,\left( \langle \bar q q\rangle^2 +
\langle \bar s s\rangle^2 \right) +
\frac{112\pi\alpha_s}{81M^4}\,\langle \bar q q\rangle \langle \bar s
s\rangle .\label{eq:SRa2}
\end{eqnarray}
We evaluate the sum rules using the input given in Table~\ref{tab:cond}.
\begin{table}[bt]
\renewcommand{\arraystretch}{1.3}
\addtolength{\arraycolsep}{3pt}
$$
\begin{array}{r@{\:=\:}l||r@{\:=\:}l}
\hline
\quark & (-0.24\pm0.01)^3\,\mbox{GeV}^3 & \squark & (1-\delta_3)\,\quark\\
\mixed & m_0^2\,\quark &  \smixed & (1-\delta_5)\mixed\\[6pt]
\displaystyle \gluon & (0.012\pm 0.006)\,
{\rm GeV}^4 & \multicolumn{2}{l}{}\\[6pt]\hline
\multicolumn{4}{c}{m_0^2 = (0.8\pm 0.1)\,{\rm GeV}^2,\quad \delta_3
  = 0.2\pm 0.2, \quad \delta_5 = 0.2\pm 0.2}\\\hline
\multicolumn{4}{c}{\overline{m}_s(2\,\mbox{GeV}) = (100\pm
20)\,\mbox{MeV}~~~\longleftrightarrow~~~ \overline{m}_s(1\,\mbox{GeV})
= (137\pm 27)\,\mbox{MeV}}\\\hline
\multicolumn{4}{c}{\alpha_s(m_Z) = 0.1187\pm 0.002  ~\longleftrightarrow~
\alpha_s(1\,\mbox{GeV}) = 0.53^{+0.06}_{-0.05}}\\\hline
\end{array}
$$
\renewcommand{\arraystretch}{1}
\addtolength{\arraycolsep}{-3pt}
\vskip-10pt
\caption[]{\sf Input parameters for sum rules at the
  renormalization scale $\mu=1\,$GeV. The value of $m_s$ is obtained
  from
  unquenched lattice calculations with $N_f=2$ flavours
as summarized in Ref.~\cite{Knechtli:2005ew}, which agrees with the results from QCD
  sum-rule calculations \cite{Narison:2005ny}. 
$\alpha_s(m_Z)$ is the PDG
  average \cite{PDG}.}\label{tab:cond}
\end{table}
The results for $f_K$ and $a_2^K$ are shown in
  Figure~\ref{fig:app1}; $f_K$ depends rather sensitively on the choice
  of $s_0$. In order to reproduce the experimental result
  $f_K=160\,$MeV, one has to choose $s_0 = 1.1\,{\rm GeV}^2$. This is
  the value we will use also for all other sum rules for $K$ matrix
  elements. For $a_2^K$, we then find
\begin{equation}\label{eq:resa1a2}
a_2^K(1\,{\rm GeV}) = 0.30\pm 0.15,
\end{equation}
which is slightly larger than the result obtained in
Ref.~\cite{BB03} and agrees with that obtained in
Ref.~\cite{KMM04}. 
For $a_1^K$, we obtain the same result as Refs.~\cite{BZ05,BZ06}:
\begin{equation}
a_1^K(1\,{\rm GeV}) = 0.06\pm 0.03.
\end{equation}

\section{Sum Rules for Twist-3 Matrix Elements}\label{app:D}
\setcounter{equation}{0}

In this appendix we estimate the parameters of the twist-3 distribution amplitudes 
$f_{3K}$, $\lambda_{3K}$ and $\omega_{3K}$ from  QCD sum rules. 
Our approach is similar to that of Ref.~\cite{CZ}, where $f_{3\pi}$
and $\omega_{3\pi}$ have been determined, and based on the
calculation of the 
correlation function of a  non-local light-ray operator, which enters the definition of the three-particle
distribution amplitude (\ref{eq:3pT3}), with the corresponding local operator:
\begin{eqnarray}
\Pi_D & = & i \int\! d^4 y \, e^{-i p y} \langle 0 | T \bar q(z) i
\sigma_{\mu z} \gamma_5 gG_{\mu  z}(vz) s(0) \bar s(y) i
\sigma_{\nu z} \gamma_5 gG_{\nu  z}(y) q(y)| 0\rangle
\nonumber\\
& \equiv & (pz)^4 \int {\cal D}\underline{\alpha}\,e^{-ipz(\alpha_2+
  v\alpha_3)} \,\pi_D(\underline{\alpha})\,.
\label{eq:PiD}
\end{eqnarray}
We also study the correlation function of that operator with the
pseudoscalar current:\footnote{Note that the
  currents in  $\Pi_{ND}$ contain no factors $i$, in
  contrast to $\Pi_D$. This is so as to obtain a positive spectral density.} 
\begin{eqnarray}
 \Pi_{ND} & = & i \int\! d^4 y\, e^{-i p y} \langle 0 | T \bar q(z) 
\sigma_{\mu z} \gamma_5 gG_{\mu  z}(vz) s(0) \bar s(y) \gamma_5 q(y)| 0\rangle
\nonumber\\
& \equiv & (pz)^2 \int {\cal D}\underline{\alpha}\,e^{-ipz(\alpha_2+
  v\alpha_3)} \,\pi^{(1)}_{ND}(\underline{\alpha})\,;
\label{eq:PiND}
\end{eqnarray}
for brevity, we do not show the Wilson lines in the non-local
operators. Our calculation goes beyond that done in Ref.~\cite{CZ} by
including $SU(3)$-breaking corrections, and by also studying sum rules based on
the non-diagonal correlation function, which allows a non-trivial
consistency check of the results.

Somewhat imprecisely, we will refer to $\Pi_D$ and  $\Pi_{ND}$ 
as ``diagonal'' and ``non-diagonal'' correlation functions, respectively.
The hadronic representation of the non-diagonal correlation function $\Pi_{ND}$ only contains
pseudoscalar $J^P = 0^-$ contributions,  whereas the diagonal
correlation function $\Pi_D$ 
also contains contributions of states with higher spin, $J^P = 2^-$ and $J^P= 1^+$.
This is not a disadvantage, since such states all have considerably
higher masses than the $K$ meson, 
and can effectively be thought of as parts of the continuum contribution. 
For reasons that will become clear below, we have also calculated a correlation function
similar to (\ref{eq:PiD}), but with operators of opposite parity:
\begin{eqnarray}
\bar \Pi_D & = & i \int\! d^4 y \, e^{-i p y} \langle 0 | T \bar q(z) 
\sigma_{\mu z}  gG_{\mu  z}(vz) s(0) \bar s(y) 
\sigma_{\nu z} gG_{\nu  z}(y) q(y)| 0\rangle
\nonumber\\
& \equiv & (pz)^4 \int {\cal D}\underline{\alpha}\,e^{-ipz(\alpha_2+
  v\alpha_3)} \,\bar \pi_D(\underline{\alpha})\,.
\label{eq:PiD2}
\end{eqnarray}
For the diagonal correlation function we find, using factorization approximation for the 
four-quark condensates and dropping terms that vanish after Borel transformation:
\begin{eqnarray}
\pi_D(\underline{\alpha}) 
&=& \frac{\alpha_s}{\pi^3}\, \alpha_1\alpha_2\alpha_3^2 \,p^2
  \ln\,\frac{\mu^2}{-p^2} 
  -\frac{2}{3}\,\frac{\alpha_s}{\pi}\,\frac{m_s \langle
  \bar s s\rangle}{p^2}\,\alpha_2\alpha_3^2\delta(\alpha_1)
  -\frac{2}{3}\,\frac{\alpha_s}{\pi}\,\frac{m_q \langle
  \bar q q\rangle}{p^2}\,\alpha_1\alpha_3^2\delta(\alpha_2)
\nonumber\\
&& {}+\frac{\alpha_s}{\pi}\,\frac{m_s \langle \bar s
  \sigma gG s\rangle}{p^4}\,\left(-\frac{7}{72}\,\alpha_3^2 +
  \frac{1}{4}\,\alpha_2\alpha_3 + \frac{1}{9}\,i (pz) \alpha_2
  \alpha_3^2 \right)\delta(\alpha_1)
\nonumber\\
&&{}+\frac{\alpha_s}{\pi}\,
\frac{m_q \langle \bar q \sigma gG
  q\rangle}{p^4}\,\left(-\frac{7}{72}\,\alpha_3^2 +
  \frac{1}{4}\,\alpha_1\alpha_3 - \frac{1}{9}\,i (pz) \alpha_1
  \alpha_3^2 \right)\delta(\alpha_2)
\nonumber\\
&&{} + \frac{\alpha_s^2\langle \bar s
  s\rangle^2}{p^4}\,\left(\frac{44}{243}\, \alpha_3^2 +\frac{2}{9}\,
  \alpha_2\alpha_3 - \frac{32}{243}\, i (pz)
  \alpha_2\alpha_3^2\right)\delta(\alpha_1)
\nonumber\\
&&{} + \frac{\alpha_s^2\langle \bar q
  q\rangle^2}{p^4}\,\left(\frac{44}{243}\, \alpha_3^2 +\frac{2}{9}\,
  \alpha_1\alpha_3 + \frac{32}{243}\, i (pz)
  \alpha_1\alpha_3^2\right)\delta(\alpha_2)
\nonumber\\
&&{}+ \frac{32}{27}\,\frac{\alpha_s^2 \langle\bar s s \rangle \langle \bar q
  q\rangle}{p^4}\,\Big(\alpha_1 \alpha_3\, \delta(\alpha_2) + \alpha_2
  \alpha_3\,\delta(\alpha_1) \Big).
\label{eq:piD}
\end{eqnarray}
To this accuracy, the expressions for $\pi_D$ and $\bar\pi_D$ are almost the same, the only difference being
that in $\bar\pi_D$ the last term in (\ref{eq:piD}), the contribution
of $\quark\squark$, comes with a minus sign. In the chiral limit, we can compare the above 
result with that obtained in Ref.~\cite{CZ}: we find agreement for the perturbative
contribution, but a different answer for the contribution of the four-quark condensates. 
 The leading-order contribution ${\mathcal O}(\alpha_s)$ of the 
gluon condensate as well as that of the dimension-6 triple-gluon
condensate $\langle g^3 f G^3\rangle$ both vanish. We also have calculated the contribution of the gluon
condensate in the local limit, $e^{-ipz(\alpha_2+v\alpha_3)}\to 1$, and find
\begin{equation}
\left.\Pi_D\right|_{\langle\frac{\alpha_s}{\pi}\,G^2\rangle} =
\left.\bar \Pi_D\right|_{\langle\frac{\alpha_s}{\pi}\,G^2\rangle} = 
-\frac{89}{5184}\,\frac{\alpha_s}{\pi}\,\left\langle\frac{\alpha_s}{\pi}\,G^2\right\rangle \,\frac{(pz)^4}{p^2},
\end{equation}
which differs from the result obtained in Ref.~\cite{CZ}. In
particular, we do not reproduce the logarithmic term quoted in \cite{CZ}.
   
{}For the non-diagonal correlation function we find
\begin{eqnarray}
\pi_{ND}(\underline{\alpha}) &=& \frac{\alpha_s}{2\pi^3}\, 
\alpha_1\alpha_2\alpha_3
\left(\frac{1}{1-\alpha_1}+\frac{1}{1-\alpha_2}
\right) p^2 \ln\,\frac{\mu^2}{-p^2}\nonumber\\
&&{}+ \frac{1}{12}\gluon \,
\frac{\alpha_1\alpha_2\delta(\alpha_3)}{\alpha_1 m_q^2+ \alpha_2 m_s^2
  - \alpha_1\alpha_2 p^2}  
\nonumber\\
&&{}
+ \frac{\alpha_s}{3\pi}\frac{1}{p^2} 
   \Big[m_q\quark \alpha_1^2\delta(\alpha_2) + m_s\squark \alpha_2^2 \delta(\alpha_1)\Big]
\nonumber\\
&&{}
+ \frac{2\alpha_s}{3\pi}\frac{1}{p^2}\,\left[\alpha_3+\alpha_3^2
\left(
     \ln\frac{\mu^2}{-p^2} - \ln(\bar \alpha_3 \alpha_3) - 1\right)\right] 
   \Big[m_s\quark \delta(\alpha_2) + m_q\squark \delta(\alpha_1)\Big]
\nonumber\\
&&{}
+\left[\frac{16}{27}\pi\alpha_s \squark^2
  +\frac{1}{6}{m_s}\smixed\right]
  \frac{1}{p^4} \delta(\alpha_1)\delta(\alpha_3)
\nonumber\\
&&{}
+\left[\frac{16}{27}\pi\alpha_s \quark^2 +\frac{1}{6}{m_q}\mixed\right]
  \frac{1}{p^4} \delta(\alpha_2)\delta(\alpha_3)\nonumber\\
&&{}
+\frac{16\pi\alpha_s}{9p^4}\quark\squark \delta(\alpha_1)\delta(\alpha_2).   
\label{eq:piND}
\end{eqnarray}

The sum rules for the couplings $f_{3K}$, $\lambda_{3K}$ and $\omega_{3K}$ are derived
by expanding the correlation functions in powers of $(pz)$:
\begin{eqnarray}
\Pi_{D} &=& (pz)^4 \left\{\Pi^{(0)}_D + i (pz) \left[ \Pi_D^{(\lambda)} + (2v-1)
\Pi_D^{(\omega)}\right] + {\mathcal O}((pz)^2)\right\}\,,
\nonumber\\
\Pi_{ND} &=& (pz)^2 \left\{\Pi^{(0)}_{ND} + i (pz) \left[ \Pi_{ND}^{(\lambda)} + (2v-1)
\Pi_{ND}^{(\omega)}\right] + {\mathcal O}((pz)^2)\right\}\,.
\label{eq:SRdiag}
\end{eqnarray}
Comparing these expressions with the corresponding expansion of the
$K$ contribution to the correlation functions expressed in terms of
the DA (\ref{eq:turk}), one obtains 
\begin{eqnarray}
  4 f_{3K}^2 e^{-m_K^2/M^2} &=& {\mathcal B}\left[\Pi^{(0)}_D\right](M^2)\,,
\nonumber\\
   \frac17\, f_{3K}^2 \lambda_{3K} e^{-m_K^2/M^2} &=& 
   {\mathcal B}\left[\Pi^{(\lambda)}_D+\frac12 \Pi^{(0)}_D \right](M^2)\,,
\nonumber\\
   -\frac{3}{14}\, f_{3K}^2 \omega_{3K} e^{-m_K^2/M^2} &=& 
   {\mathcal B}\left[\Pi^{(\omega)}_D+\frac{3}{14} \Pi^{(0)}_D\right](M^2)\,,
\end{eqnarray}
and similarly
\begin{eqnarray}
 2 f_{3K}\frac{f_K m_K^2}{m_s+m_q} e^{-m_K^2/M^2} &=& 
{\mathcal B}\left[\Pi^{(0)}_{ND}\right](M^2)\,,
\nonumber\\
 \frac{1}{14}\, f_{3K} \lambda_{3K} \,\frac{f_K m_K^2}{m_s+m_q}\,
  e^{-m_K^2/M^2} &=& 
{\mathcal B}\left[\Pi^{(\lambda)}_{ND}+\frac12 \Pi^{(0)}_{ND}\right](M^2)\,,
\nonumber\\
 -\frac{3}{28}\,\omega_{3K} f_{3K}\,\frac{f_K m_K^2}{m_s+m_q} \,
  e^{-m_K^2/M^2} &=& 
{\mathcal B}\left[\Pi^{(\omega)}_{ND}+\frac{3}{14}\Pi^{(0)}_{ND}\right](M^2)\,,
\label{eq:SRnondiag}
\end{eqnarray}
from the diagonal and non-diagonal correlation functions, respectively.
Here and below ${\mathcal B[\ldots](M^2)}$ stands for the Borel transformation with respect to $p^2$;
$M^2$ is the Borel parameter. 

From $\Pi_D$, we obtain the following sum rule for $f_{3K}$:
\begin{eqnarray}
4 \left.f_{3K}^2\right|_D e^{-m_K^2/M^2} & = & \frac{\alpha_s}{360\pi^3} \int_0^{s_0}
ds s e^{-s/M^2} + \frac{\alpha_s}{18\pi}\left( m_s\squark +
m_q\quark\right)\nonumber\\
&&{}+\frac{89}{5184}\,\frac{\alpha_s}{\pi}\,\left\langle\frac{\alpha_s}{\pi}\,G^2\right\rangle +
\frac{\alpha_s}{108\pi}\,\frac{1}{M^2}\left( m_s\smixed +
m_q\mixed\right) \nonumber\\
&&{}+\frac{71}{729}\,\frac{\alpha_s^2}{M^2}\,\left(\quark^2+\squark^2\right)
+ \frac{32}{81}\, \frac{\alpha_s^2}{M^2}\,  \quark \squark\,,\label{f3-SR1}
\end{eqnarray}
where the subscript $D$ indicates that this sum rule is derived from
the correlation function $\Pi_D$. The last term on the right-hand side
comes from the factorisation of  the four-quark condensate  $(\bar q \sigma_{\mu\nu}t^A q)(\bar q
\sigma_{\mu\nu}t^A q)$.
In Ref.~\cite{CZ}, the authors have argued that this term, which
induces a large power correction in their sum rule for $f_{3\pi}$, is
unreliable because of a potential breakdown of the factorisation
approximation for that particular condensate; they suggested to determine $f_{3\pi}$
from a sum rule derived from the sum of the correlation functions $\Pi_D +
\bar\Pi_D$ instead, where these large contributions cancel.
Indeed, the Dirac structures $\sigma_{\mu\nu}$ and
$i\sigma_{\mu\nu}\gamma_5$ are not independent, but related by $i\sigma_{\mu\nu}\gamma_5=
-\frac{1}{2}\,\epsilon_{\mu\nu\rho\sigma}\sigma_{\rho\sigma}$, which
induces the relation
\begin{eqnarray*}
\bar\Pi_D &=&i \int\! d^4 y \, e^{-i p y} \langle 0 | T \bar q(0) 
i\sigma_{\mu z}\gamma_5  gG_{\nu  z}(0) s(0) \bar s(y) 
i\sigma_{\mu z}\gamma_5 gG_{\nu  z}(y) q(y)| 0\rangle\\
&&{} - i \int\! d^4 y
\, e^{-i p y} \langle 0 | T \bar q(0) 
i\sigma_{\mu z}\gamma_5  gG_{\nu  z}(0) s(0) \bar s(y) 
i\sigma_{\nu z}\gamma_5 gG_{\mu  z}(y) q(y)| 0\rangle\,.
\end{eqnarray*}
$\bar\Pi_D$ receives no contributions from $0^-$ states 
because their contributions to the two correlation functions on the
right-hand side are equal and cancel in the difference; the same
applies to $1^+$ states, so that the
lowest resonance contributing to $\bar\Pi_D$ is $1^-$. These states
can safely be included in the continuum so that
it is possible to extract $f_{3K}$ from the sum of correlation
functions $\Pi_D+\bar\Pi_D$. On the other hand, our sum rule
(\ref{f3-SR1}), derived from $\Pi_D$ only, with the correct coefficients for gluon and four-quark
condensates, is actually not very sensitive to the term in $\quark \squark$, but
dominated by the gluon condensate. As there is no strong theoretical
argument in favour or disfavour of either diagonal sum rule, the one based
on $\Pi_D$ and the one based on $\Pi_D+\bar\Pi_D$, we decide to
use both. We also determine $f_{3K}$ from a  third sum rule  based on the non-diagonal
correlation function $\Pi_{ND}$; the difference between these three results
will be interpreted as theoretical uncertainty.

Explicitly, we obtain, in addition to (\ref{f3-SR1}), the following
sum rules for $f_{3K}$, with the index indicating the underlying
correlation function:
\begin{eqnarray}
4 \left.f_{3K}^2\right|_{D+\bar D} e^{-m_K^2/M^2} & = & \frac{\alpha_s}{180\pi^3} \int_0^{s_0}
ds s e^{-s/M^2} + \frac{\alpha_s}{9\pi}\left( m_s\squark +
m_q\quark\right)\nonumber\\
&&{}+\frac{89}{2592}\,\frac{\alpha_s}{\pi}\,\left\langle\frac{\alpha_s}{\pi}\,G^2\right\rangle +
\frac{\alpha_s}{54\pi}\,\frac{1}{M^2}\left( m_s\smixed +
m_q\mixed\right) \nonumber\\
&&{}+\frac{142}{729}\,\frac{\alpha_s^2}{M^2}\,\left(\quark^2+\squark^2\right),\label{f3-SR2}
\end{eqnarray}
\begin{eqnarray}
2\lefteqn{ \left.f_{3K}\right|_{ND}\, \frac{f_Km_K^2}{m_s+m_q} \,e^{-m_K^2/M^2}  = 
\frac{\alpha_s}{72\pi^3} \int_0^{s_0} ds s e^{-s/M^2} +
\frac{1}{12}\,\left\langle\frac{\alpha_s}{\pi}\,G^2\right\rangle-\frac{\alpha_s}{9\pi}\,(m_q\quark
+ m_s\squark)} \hspace*{3.4cm}\nonumber\\
&&{}-\frac{2}{9}\,\frac{\alpha_s}{\pi}(m_s\quark + m_q\squark)\left( \frac{8}{3}+\gamma_E -
\ln\,\frac{M^2}{\mu^2} + \int_{s_0}^\infty
\frac{ds}{s}\,e^{-s/M^2}\right)\nonumber\\
&&{}+\frac{1}{6M^2}(m_s\smixed + m_q\mixed)\nonumber\\
&&{} +
\frac{16}{27}\,\frac{\pi\alpha_s}{M^2}\,(\quark^2+\squark^2) +
\frac{16}{9}\,\frac{\pi\alpha_s}{M^2}\,\quark\squark\,.\label{f3-SR3}
\end{eqnarray}
The sum rules for $f_{3\pi}$ are obtained by taking the chiral limit of
the above expressions.

As for $\omega_{3K}$, we have not calculated the gluon-condensate contribution to the
diagonal sum rule, which is expected to be dominant,
so we cannot use the diagonal sum rule and only consider the
non-diagonal one:
\begin{eqnarray}
\lefteqn{2 \left.(f_{3K} \omega_{3K})\right|_{ND}\, \frac{ f_Km_K^2}{m_s+m_q} \,e^{-m_K^2/M^2} = 
-\frac{\alpha_s}{60\pi^3} \int_0^{s_0} ds s e^{-s/M^2}
+\frac{5}{27}\,\frac{\alpha_s}{\pi}\,(m_q\quark + m_s\squark)}\hspace*{2.5cm} \nonumber\\
& - &\frac{2}{3}\,\frac{\alpha_s}{\pi}(m_s\quark + m_q\squark)\left( \frac{8}{3}+\gamma_E -
\ln\,\frac{M^2}{\mu^2} + \int_{s_0}^\infty
\frac{ds}{s}\,e^{-s/M^2}\right)\nonumber\\
&-&\frac{1}{3}\,\left\langle\frac{\alpha_s}{\pi}\,G^2\right\rangle+\frac{2}{3M^2}(m_s\smixed
+ m_q\mixed)\nonumber\\
& -&
\frac{64}{27}\,\frac{\pi\alpha_s}{M^2}\,(\quark^2+\squark^2) +
\frac{256}{27}\,\frac{\pi\alpha_s}{M^2}\,\quark\squark\,.\hspace*{1cm}\label{w3-SR}
\end{eqnarray}
In evaluating this sum rule, we replace $f_{3K}$ by the expression
obtained from (\ref{f3-SR3}).

As for $\lambda_{3K}$, the gluon-condensate contribution is suppressed by
a factor $m_s^2-m_q^2$ by virtue of G-parity and can safely be neglected
in the diagonal sum rule. We did calculate this contribution for the non-diagonal
sum rule, though, where indeed it gives only a small contribution. On the
other hand, the $\quark\squark$ contribution is also absent because of 
G-parity, so that the two diagonal sum rules for
$\left.f_{3K}^2\lambda_{3K}\right|_{D}$ and
$\left.f_{3K}^2\lambda_{3K}\right|_{D+\bar D}$ differ by a global factor 2. As the
values of $\left.f_{3K}^2\right|_D$ and $\left.f_{3K}^2\right|_{D+\bar
  D}$ also differ by a factor of  approximately 2, this theoretical
uncertainty cancels to a large extent. The sum rules read:
\begin{eqnarray}
\lefteqn{4 \left.(f_{3K}^2 \lambda_{3K})\right|_{D} e^{-m_K^2/M^2} =  
-\frac{14}{45}\,\frac{\alpha_s}{\pi}\left( m_s\squark -
m_q\quark\right)}\hspace*{1.5cm} \nonumber\\
&&{} + \frac{35}{512}\,\frac{\alpha_s}{\pi}\,\frac{1}{M^2}\left( m_s\smixed -
m_q\mixed\right)+\frac{7}{9}\,\frac{\alpha_s^2}{M^2}\,\left(\quark^2-\squark^2\right),\label{l3-SR1}
\end{eqnarray}
\begin{eqnarray}
\lefteqn{
4 \left.(f_{3K}^2 \lambda_{3K})\right|_{D+\bar D} e^{-m_K^2/M^2} =  
-\frac{28}{45}\,\frac{\alpha_s}{\pi}\left( m_s\squark -
m_q\quark\right)}\hspace*{1.5cm} \nonumber\\
&&{} + \frac{35}{216}\,\frac{\alpha_s}{\pi}\,\frac{1}{M^2}\left( m_s\smixed -
m_q\mixed\right)+\frac{14}{9}\,\frac{\alpha_s^2}{M^2}\,\left(\quark^2-\squark^2\right),\label{l3-SR2}
\end{eqnarray}
\begin{eqnarray}
\lefteqn{2 \left.(f_{3K} \lambda_{3K})\right|_{ND}\, \frac{f_Km_K^2}{m_s+m_q} \,e^{-m_K^2/M^2}  =
\frac{7}{6}\,\frac{\alpha_s}{\pi}\,( m_s\squark-m_q\quark)}\hspace*{4cm} \nonumber\\
&-&\frac{7}{9}\,\frac{\alpha_s}{\pi}(m_s\quark - m_q\squark)\left( \frac{8}{3}+\gamma_E -
\ln\,\frac{M^2}{\mu^2} + \int_{s_0}^\infty
\frac{ds}{s}\,e^{-s/M^2}\right)\nonumber\\
&-&{}\frac{7}{6M^2}\,\left\langle\frac{\alpha_s}{\pi}\,G^2\right\rangle
(m_s^2-m_q^2)\left(1+\gamma_E-\ln\frac{M^2}{\mu^2} -
M^2\int_{s_0}^\infty 
\frac{ds}{s^2}\,e^{-s/M^2}\right)\nonumber\\
&+&\frac{7}{3M^2}(m_q\mixed-m_s\smixed) +
\frac{224}{27}\,\frac{\pi\alpha_s}{M^2}\,(\quark^2-\squark^2)\,.\nonumber\\[-15pt]\label{l3-SR3}
\end{eqnarray}
Again, when evaluating these sum rules, we replace $f_{3K}$ by the
corresponding expressions obtained from (\ref{f3-SR1}), (\ref{f3-SR2}),
and (\ref{f3-SR3}).
 
\begin{figure}
$$\epsfxsize=0.46\textwidth\epsffile{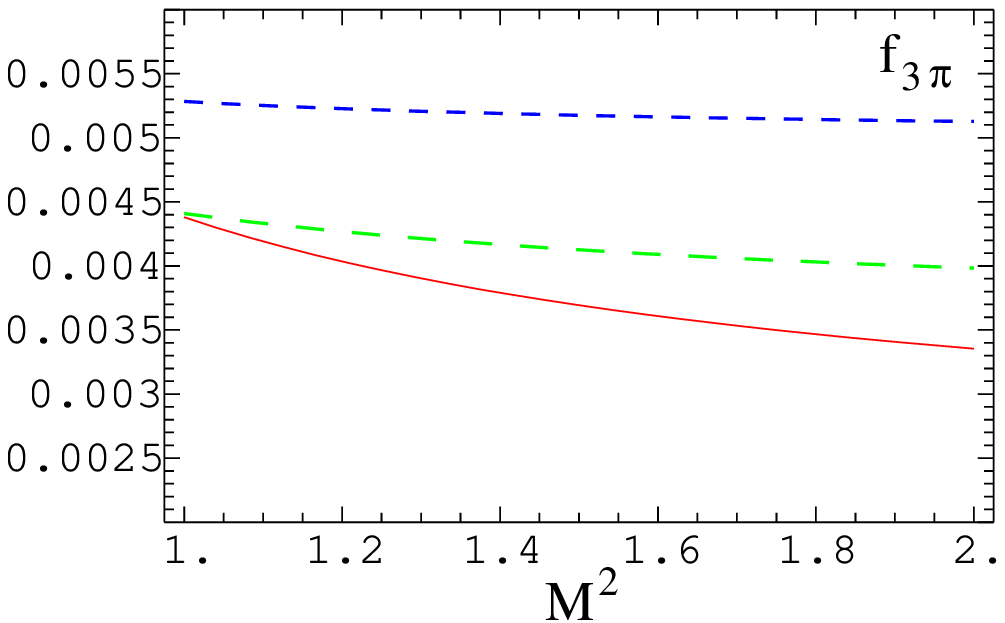}\qquad
\epsfxsize=0.46\textwidth\epsffile{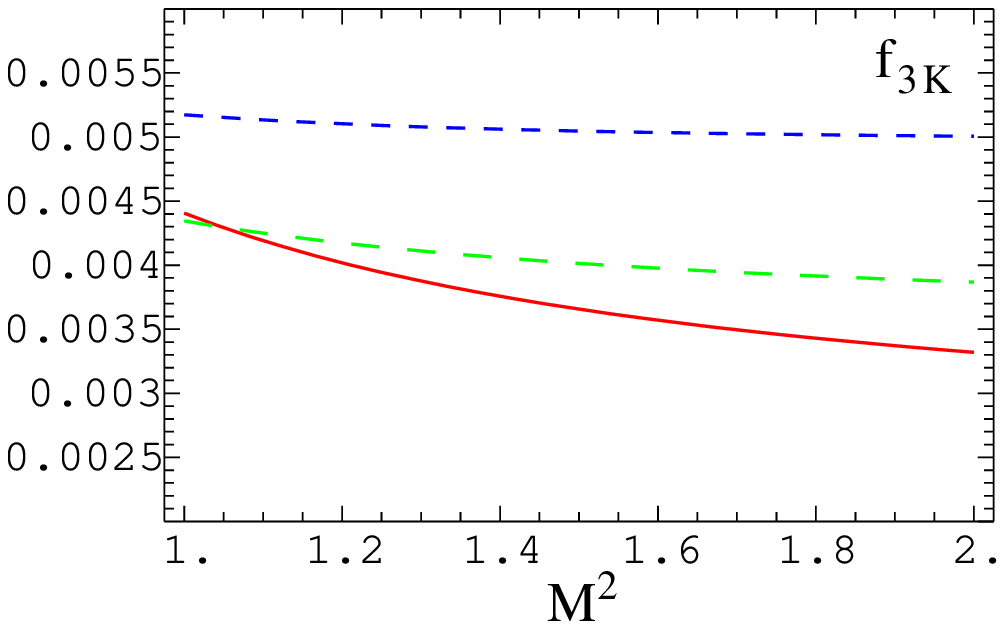}$$
\vspace*{-20pt}
\caption[]{\sf Left panel: $f_{3\pi}$ as a function of the Borel
  parameter, calculated from the non-diagonal sum rule (\ref{f3-SR3}) (red, solid
  line), the pure-parity diagonal sum rule (\ref{f3-SR1}) (green, long dashes) and
  the mixed-parity diagonal sum rule (\ref{f3-SR2}) (blue, short dashes); $s_0 =
  0.8\,{\rm GeV}^2$. Right
  panel: same for $f_{3K}$; $s_0 = 1.1\,{\rm GeV}^2$.}\label{fig:f3}
$$\epsfxsize=0.46\textwidth\epsffile{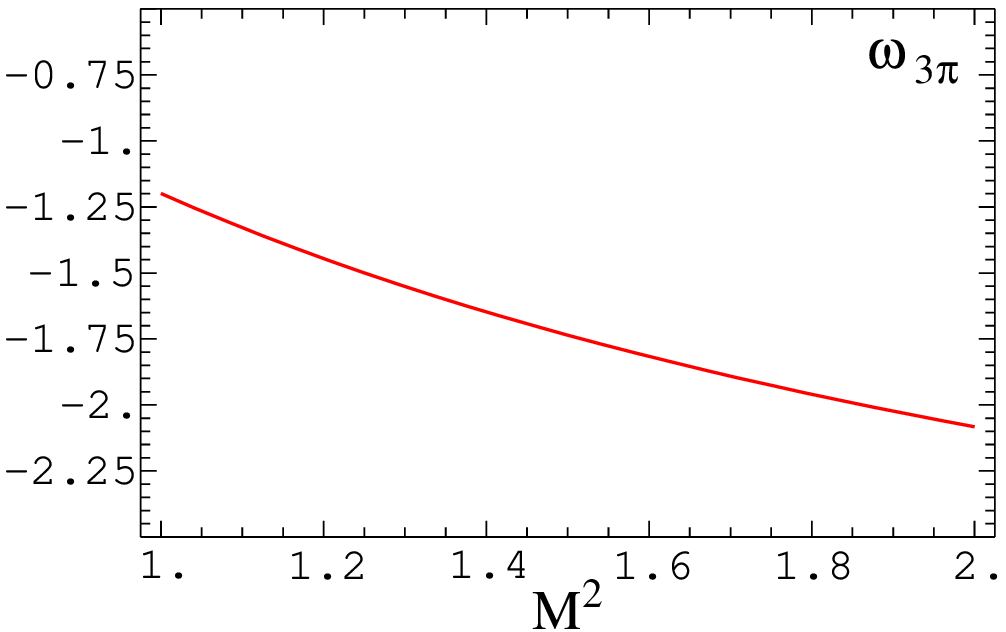}\qquad
\epsfxsize=0.46\textwidth\epsffile{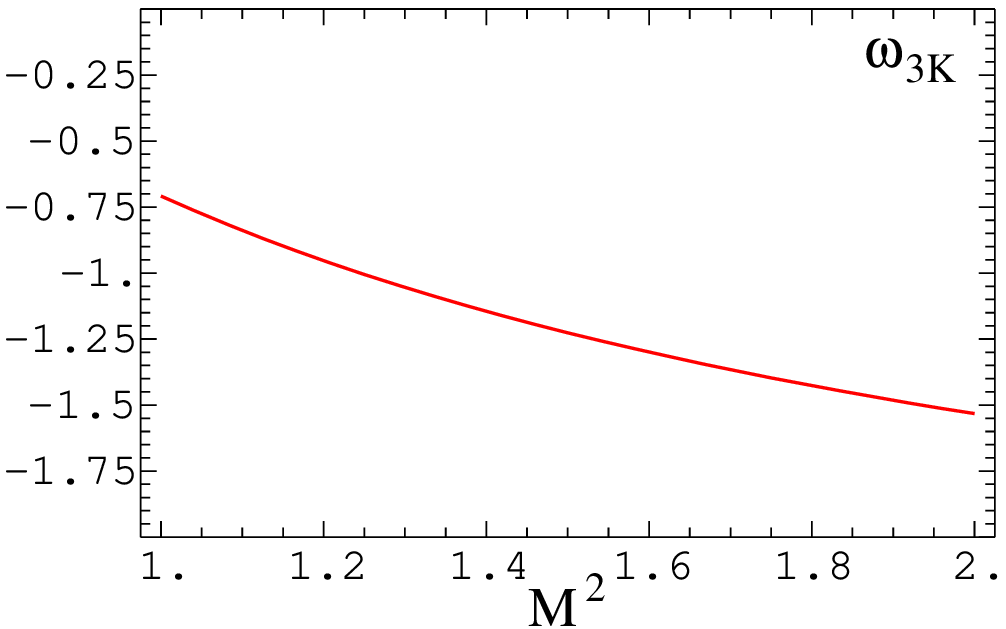}$$
\vspace*{-20pt}
\caption[]{\sf Left panel: $\omega_{3\pi}$ as a function of the Borel
  parameter from the non-diagonal sum rule (\ref{w3-SR}); $s_0 =
  0.8\,{\rm GeV}^2$. Right
  panel: same for $\omega_{3K}$; $s_0 = 1.1\,{\rm GeV}^2$. The results
  from diagonal sum rules are not shown because the gluon-condensate
  contribution is unknown.}\label{fig:w3}
$$\epsfxsize=0.46\textwidth\epsffile{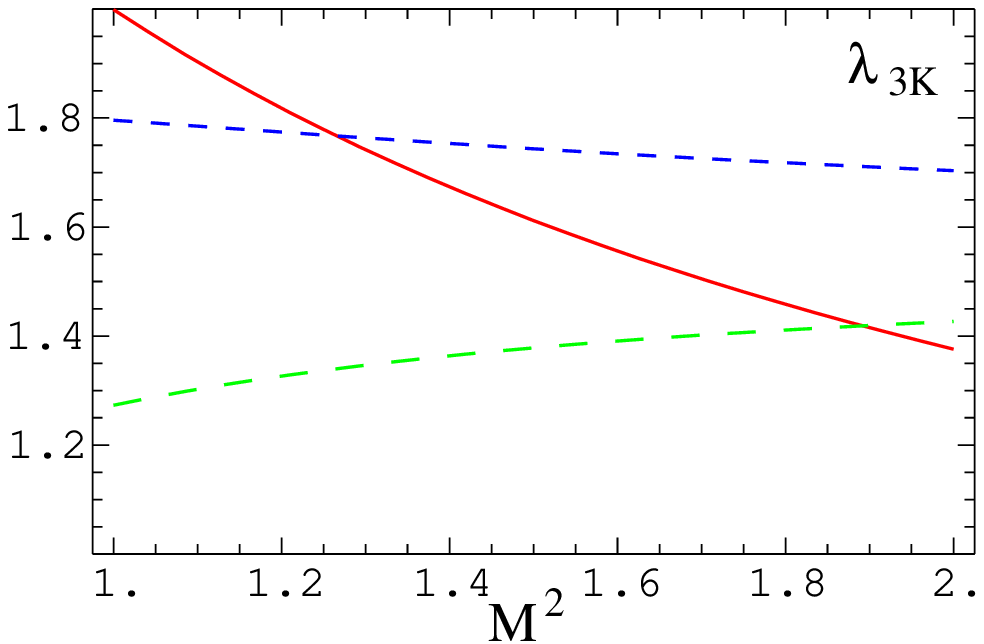}$$
\vspace*{-20pt}
\caption[]{\sf $\lambda_{3K}$ as a function of the Borel
  parameter, calculated from the non-diagonal sum rule (\ref{l3-SR3}) (red, solid
  line), the pure-parity diagonal sum rule (\ref{l3-SR1}) (green, long dashes) and
  the mixed-parity diagonal sum rule (\ref{l3-SR2}) (blue, short dashes); $s_0 =
  1.1\,{\rm GeV}^2$; $\lambda_{3\pi}=0$ by virtue of G-parity.
}\label{fig:l3}
\end{figure}

The numerical results from all these sum rules are shown in
Figures~\ref{fig:f3} to \ref{fig:l3}. As for $f_{3\pi}$ and $f_{3K}$,
all three sum rules yield very similar results, which is a strong
indication for the consistency of the approach. The diagonal sum
rules are very stable in $M^2$, the non-diagonal ones less so. Taking into account
the uncertainties of the input parameters as given in
Table~\ref{tab:cond} and the difference in the results from the
different sum rules, we obtain the estimates
\begin{equation}
f_{3\pi}(1\,{\rm GeV}) = (0.0045\pm 0.0015)\,{\rm GeV}^2,\quad
f_{3K}(1\,{\rm GeV}) = (0.0045\pm 0.0015)\,{\rm GeV}^2.
\end{equation}
The effect of $SU(3)$ breaking is very small, 
\begin{equation}
  f_{3K}/f_{3\pi} = 0.98\pm 0.03\,,
\end{equation}
as all sum rules are
dominated by the contribution of the gluon condensate. Note that our
value for $f_{3\pi}$ is about 50\% larger than the one obtained in Ref.~\cite{CZ},
which is due to, as we believe, the incorrect results for the contributions of the
gluon and  four-quark condensate contributions obtained in this paper.
As for $\omega_{3}$, as explained above, we only evaluate the non-diagonal
sum rule. We find that the
sum rules are less stable in $M^2$, as with $f_3$, and that now the
effect of $SU(3)$ breaking is more prominent. Our final estimate is
\begin{equation}
\omega_{3\pi}(1\,{\rm GeV}) = -1.5\pm 0.7\,,\qquad \omega_{3K}(1\,{\rm GeV}) = -1.2\pm 0.7\,,
\end{equation}
where the error reflects in particular the uncertainty of the value of
the gluon condensate. Our result is to be compared with that of
Ref.~\cite{CZ}, $\omega_{3\pi} \approx -3$.
Finally, $\lambda_{3K}$ can be determined from
three sum rules, as the gluon-condensate contribution is
suppressed by a factor $m_s^2$. All three sum
rules yield perfectly consistent values, despite the fact that the two
diagonal sum rules (\ref{l3-SR1}) and (\ref{l3-SR2}) differ by an overall
factor of 2, which, as expected, is largely cancelled by the different values of
$f_{3K}^2|_D$ and $f_{3K}^2|_{D+\bar D}$. We obtain
\begin{equation}
\lambda_{3K}(1\,{\rm GeV}) = 1.6\pm 0.4;
\end{equation}
the error is smaller than for $\omega_{3K}$ because  the gluon
condensate is suppressed. This result is new.

\section{Sum Rules for Twist-4 Matrix Elements}\label{app:E}
\setcounter{equation}{0}

The aim of this section is to estimate the decay constant $\delta_K^2$ that 
determines the normalization of twist-4 distribution amplitudes. To this end we define 
the currents 
\begin{eqnarray}
   J_\mu^A &=& \bar q \,g\wt G_{\mu\alpha}\gamma_\alpha s\,, \qquad
   J_\mu^V \, = \,  \bar q \, g\wt G_{\mu\alpha}\gamma_\alpha\gamma_5 s\,, 
\label{4-1}
\end{eqnarray}
with quantum numbers $J^P = 1^+$ and $1^-$, respectively,
and calculate the correlation functions 
\begin{eqnarray}
\Pi_{\mu\nu}^{A,V} &=& i\int\! d^4 x\, e^{ipx}\,
  \bra 0| T J_\mu^{A,V}(x)  (J_\nu^{A,V})^\dagger(0)|0\ket\, 
    = \, p_\mu p_\nu \,\Pi_{0}^{A,V}(p^2) -g_{\mu\nu} \, \Pi_{1}^{A,V}(p^2)\,, 
\label{4-2}
\end{eqnarray}
taking into account contributions of operators with dimension up to eight. 
Note that the relative sign between $f_K$ and $\delta_K^2$ can be fixed from the 
non-diagonal correlation function of $J^A_\mu$ and the axial vector current. This calculation 
was done in Ref.~\cite{misuse} and will not be repeated here; the result is that  
$\delta^2_K$ is positive.   

Similar correlation functions have been considered in the past, 
mainly in connection with searches for exotic
quark--antiquark--gluon mesons \cite{exotic}. We obtain
\begin{eqnarray}
 \Pi_{0}^{A,V} &=& \frac{\alpha_s}{160 \pi^3} p^4 \ln \frac{\mu^2}{-p^2} +
                 \frac{1}{72} \gluon \ln \frac{\mu^2}{-p^2}
\nonumber\\
&&{}+  \frac{\alpha_s}{6\pi}\big[m_q \quark + m_s \squark\big] \ln \frac{\mu^2}{-p^2}
    \mp  \frac{2\alpha_s}{9\pi}\big[m_s \quark + m_q \squark\big] \ln \frac{\mu^2}{-p^2}
\nonumber\\
&&{}\mp \frac{8\pi\alpha_s}{9p^2} \quark\squark + 0\cdot \bra g^3 f G^3\ket
\nonumber\\
&&{} + \frac{5}{108}
       \frac{\alpha_s}{\pi}\frac{1}{p^2} \big[m_q \mixed + m_s \smixed\big] 
\nonumber\\
&&{} \pm 
\Big[\frac{1}{9} \ln\frac{\mu^2}{-p^2} +\frac{2}{27}\Big]
\frac{\alpha_s}{\pi}\frac{1}{p^2} \big[m_s \mixed + m_q \smixed\big]
\nonumber\\
&&{}   -\frac{25\pi\alpha_s}{324p^4} m_0^2\big[\quark^2 + \squark^2\big]
    \pm \frac{143\pi\alpha_s}{162p^4} m_0^2 \quark\squark
\nonumber\\
&&{} + \frac{\pi}{18p^4}\gluon \big[m_q \quark + m_s \squark\big]\,,         
\label{4-Pi0x}\\                 
 \Pi_{1}^{A,V} &=& \frac{\alpha_s}{240 \pi^3} p^6 \ln \frac{\mu^2}{-p^2} -
                 \frac{1}{36} \gluon p^2\ln \frac{\mu^2}{-p^2}
\nonumber\\
&&{}+  \frac{\alpha_s}{6\pi}\big[m_q \quark + m_s \squark\big] p^2\ln \frac{\mu^2}{-p^2}
    \mp  \frac{\alpha_s}{18\pi}\big[m_s \quark + m_q \squark\big] p^2\ln \frac{\mu^2}{-p^2}
\nonumber\\
&&{}\mp \frac{8\pi\alpha_s}{9} \quark\squark - \frac{1}{192\pi^2}\cdot \bra g^3 f G^3\ket
\nonumber\\
&&{} -\frac{19}{144}
       \frac{\alpha_s}{\pi} \big[m_q \mixed + m_s \smixed\big]\ln \frac{\mu^2}{-p^2} 
\nonumber\\
&&{} \pm \frac{19}{144}
\frac{\alpha_s}{\pi}\big[m_s \mixed + m_q \smixed\big]\ln \frac{\mu^2}{-p^2}
\nonumber\\
&&{}   +\frac{25\pi\alpha_s}{162p^2} m_0^2\big[\quark^2 + \squark^2\big]
    \pm \frac{181\pi\alpha_s}{162p^2} m_0^2 \quark\squark
\nonumber\\
&&{} + \frac{\pi}{18p^2}\gluon \big[m_q \quark + m_s \squark\big]
     \mp \frac{\pi}{6p^2}\gluon \big[m_s \quark + m_q \squark\big]\,.         
\label{4-Pi0}                 
\end{eqnarray} 
In both cases the upper sign refers to the axial and the lower sign to
the vector correlation function,
respectively; $\Pi^A_0$ has been calculated, in the chiral limit, in
Ref.~\cite{misuse}. The quark mass corrections and the expression for
$\Pi^{A}_1$ are new.

In this work we follow the procedure proposed in Ref.~\cite{misuse} and 
write the sum rule directly for the correlation function
$\Pi_0^A$:
\begin{equation}
   f_{K}^2 \delta_K^4 e^{-m_K^2/M^2} = {\cal B}[\Pi_0^A](M^2)\,.
\label{4-SR1}
\end{equation}

\begin{figure}
$$\epsfxsize=0.47\textwidth\epsffile{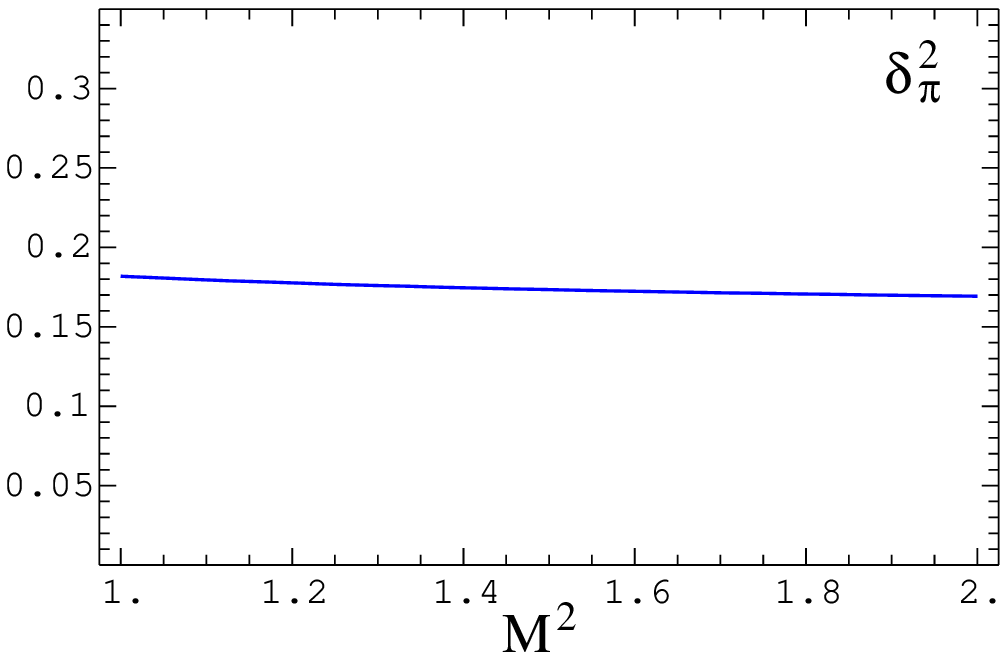}\quad
\epsfxsize=0.47\textwidth\epsffile{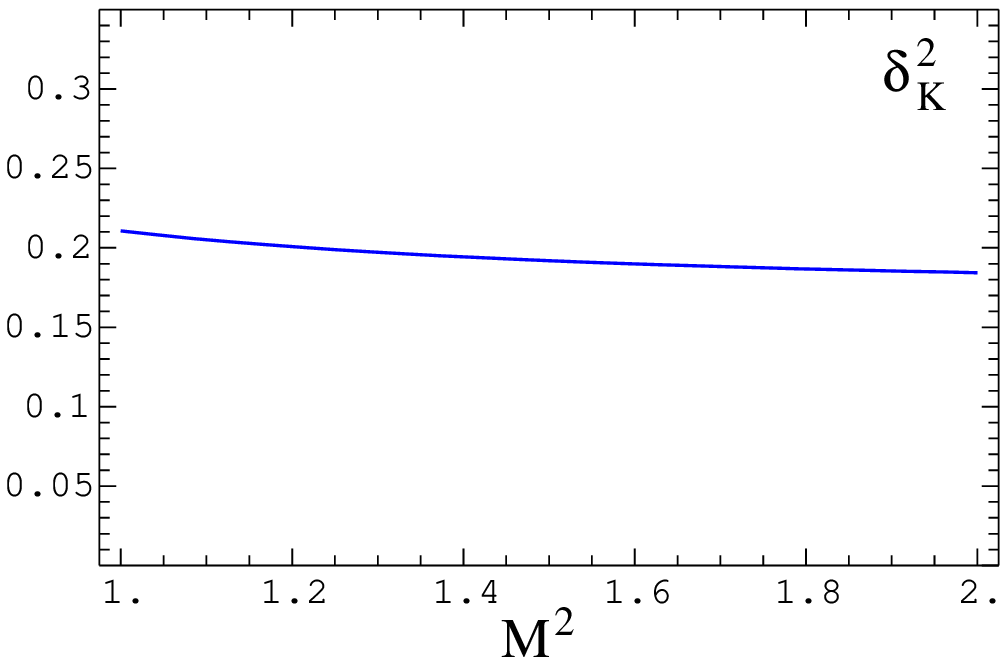}$$
\vspace*{-15pt}
\caption[]{\sf Left panel: $\delta_\pi^2$ as a function of the Borel
  parameter from the sum rule (\ref{4-SR1}); $s_0 = 0.8\,{\rm GeV}^2$. Right panel: the same for
  $\delta_K^2$; $s_0 = 1.1\,{\rm GeV}^2$.}\label{fig:d2}
\end{figure}

The results for $\delta_\pi^2$ and $\delta_K^2$ are shown in
Figure~\ref{fig:d2}. We find
\begin{equation}
\delta_\pi^2 = (0.18\pm 0.06)\,{\rm GeV}^2,\qquad \delta_K^2 = (0.20\pm 0.06)\,{\rm GeV}^2
\end{equation}
and
\begin{equation}
   \delta_K^2/\delta_\pi^2 = 1.10\pm 0.05\,,
\end{equation}
which can be compared to the estimate $(f_K\delta_K^2)/(f_\pi\delta_\pi^2)=1.07^{+0.14}_{-0.13}$
obtained in Ref.~\cite{KMM03}.

\end{document}